\documentstyle[astrobib,psfig]{mn-ab}
\newcommand{\lcdm} {\mbox{$\Lambda$CDM}}

\newcommand{\omegagas}{$\Omega_{\rm cold}$}
\newcommand{\HI}{H$_{\rm I}$}
\newcommand{\hkpc}{\mbox{$h^{-1}$ kpc}}

\newcommand{\hmsun}{\mbox{$h^{-1}$ $M_{\odot}$}}

\newcommand{\kms}{\mbox{km s$^{-1}$}}

\newcommand{\msun}{\mbox{$M_{\odot}$}}

\def\la{\mathrel{\hbox{\rlap{\hbox{\lower4pt\hbox{$\sim$}}}\hbox{$<$}}}}
\def\ga{\mathrel{\hbox{\rlap{\hbox{\lower4pt\hbox{$\sim$}}}\hbox{$>$}}}}

\title[The Nature of High-Redshift Galaxies] 
{The Nature of High-Redshift Galaxies}
\author[R. S. Somerville, J. R. Primack \& S. M. Faber]
       {Rachel S. Somerville$^{1,2}$,  Joel R. Primack$^2$
        \& S. M. Faber$^3$ \\
        $^1$Institute of Astronomy, University of Cambridge, UK\\
        $^2$Department of Physics, University of California, Santa Cruz \\
        $^3$UCO/Lick Observatory, University of California, Santa Cruz}
\begin{document}

\maketitle

\begin{abstract}
Using semi-analytic models of galaxy formation set within the Cold
Dark Matter (CDM) merging hierarchy, we investigate several scenarios
for the nature of the high-redshift ($z \ga 2$) Lyman-break galaxies
(LBGs).  We consider a ``collisional starburst'' model in which bursts
of star formation are triggered by galaxy-galaxy mergers, and find
that a significant fraction of LBGs are predicted to be
starbursts. This model reproduces the observed comoving number density
of bright LBGs as a function of redshift and the observed luminosity
function at $z\sim 3$ and $z\sim 4$, with a reasonable amount of dust
extinction.  Model galaxies at $z \sim 3$ have star formation rates,
half-light radii, I$-$K colours, and internal velocity dispersions
that are in good agreement with the data.  Global quantities such as
the star formation rate density and cold gas and metal content of the
Universe as a function of redshift also agree well. Two ``quiescent''
models without starbursts are also investigated.  In one, the star
formation efficiency in galaxies remains constant with redshift, while
in the other, it scales inversely with disc dynamical time, and thus
increases rapidly with redshift.  The first quiescent model is
strongly ruled out as it does not produce enough high redshift
galaxies once realistic dust extinction is accounted for. The second
quiescent model fits marginally, but underproduces cold gas and very
bright galaxies at high redshift.  A general conclusion is that star
formation at high redshift must be more efficient than locally. The
collisional starburst model appears to accomplish this naturally
without violating other observational constraints.
\end{abstract}


\section{Introduction}
\label{sec:intro}
With the dramatic recent advances in observational astronomy, more and
more pieces of the puzzle of galaxy formation and evolution are
becoming available. Some of the important puzzle pieces include the
number densities, colours, sizes, morphologies, internal velocity
dispersions and star formation rates of bright star-forming galaxies
spanning a redshift range from $z\sim0$ to $z\sim6$, and the
complementary information on the neutral hydrogen and metal content of
the Universe to $z\sim4$ obtained from quasar absorption systems.
However, it still remains to fit these pieces together into a
comprehensive and compelling theoretical framework.

Our window onto the high redshift ($z\ga 2$) Universe has been
expanded tremendously by the ``Lyman-break'' photometric selection
technique developed by Steidel and collaborators
\cite{steidel:92,steidel:93,steidel:96a}. This technique uses
specially developed filters to exploit the redshifted Lyman-limit
discontinuity in order to identify high-redshift candidates. Similar
techniques were exploited by
\citeN{madau:96} to identify high-redshift candidates in the Hubble
Deep Field (HDF) \cite{hdf}. Extensive spectroscopic follow-up work
at the Keck telescope has verified the accuracy of the photometric
selection technique
\cite{steidel:96b,lowenthal:97,adelberger:98,steidel:99}. The
morphologies and sizes of these objects have been studied using the
HDF sample \cite{giavalisco:96,lowenthal:97}, and their clustering
properties have been measured using the growing sample of hundreds
(now approaching a thousand) of Lyman-break galaxies (LBGs) with
spectroscopically confirmed redshifts
\cite{steidel:spike,giavalisco:98,adelberger:98}. Similar techniques
have been used to identify galaxies at $\bar{z}\sim4$
\cite{steidel:99}, and a handful of objects have been discovered with
confirmed redshifts $z \ga 5$ \cite{stern:99}.

Soon after the discovery of Lyman-break
galaxies, various interpretations of them were proposed. A view 
put forward by \citeN{steidel:96a} and
reiterated by \citeN{giavalisco:96},
\citeN{steidel:spike}, and \citeN{adelberger:98}, 
is that LBGs are located in the 
centres of massive dark matter haloes
($M \ga 10^{12} \msun$) and have been forming stars at moderate rates
over a fairly long time scale ($\ga 1$ Gyr). This scenario supposes
that LBGs are the direct
progenitors of today's massive, luminous
ellipticals and spheroids and that such objects had almost completely
assembled and begun forming stars at moderate and fairly constant
rates by $z \sim 3$.
\citeN{lowenthal:97} also equated LBGs with 
today's massive spheroids but suggested instead
that LBGs are still actively assembling and
that many exist in small-mass haloes 
raised briefly above the detection limit
by short, intense bursts of star formation.  
\citeN{trager:97} further proposed that these
small objects would later be tidally destroyed as they fell into the
potential well of larger galaxies to produce a Population II stellar
halo like that in the Milky Way.

The impressive body of observations at high redshift would seem to
offer a tantalizing promise to provide important constraints on
theories of galaxy formation. As usual, the difficulty lies in making
a connection between the visible but ill-understood portions of
galaxies and the dark matter haloes that are hidden from view but well
modelled within the Cold Dark Matter (CDM) framework.  Building on the
above massive/early-assembly picture for LBGs, several workers
investigated a model in which each LBG is associated with a dark
matter halo and luminosity scales tightly with halo mass
\cite{mf:96,adelberger:98,jing:98,wechsler:98}. This work has
demonstrated that, in such a picture, the observed number density and
clustering of LBGs at $z\sim3$ can be simultaneously accounted for if
there is approximately one galaxy per massive halo ($M \ga 10^{12}
\msun$, where this mass depends somewhat on the cosmological model).
We shall refer to this class of simple models as ``massive halo''
models. It arises quite generically from such models that LBGs must be
substantially more clustered than the underlying dark matter --- in
short, bright galaxies at high redshift are highly \emph{biased}. This
prediction has been emphasized as a major success of the massive halo
picture (e.g., \citeNP{adelberger:98}).

However, the sharp lower mass cutoff and tight link between halo mass
and luminosity in massive halo models are doubtless too simplistic.  A
step forward in realism is provided by semi-analytic models, in which
the infall rate of cooling gas determines the star formation rate, and
hence the luminosity.  Both are modelled as a function of halo mass,
which grows via hierarchical clustering, accompanied by a schematic
law for gas cooling as a function of halo gas density.
\citeN[hereafter BCFL]{bcfl} showed that semi-analytic models 
are broadly consistent with the simple massive halo picture, and that
they also match the observed abundance and clustering of LBGs at
$z\sim3$ (see also \citeNP{governato:98}).  

While these results are encouraging, there is an important caveat.
Agreement with the observed number densities of LBGs is achieved in
the above models only if internal dust extinction in LBGs is ignored.
Much evidence has now accumulated to the effect that dust in LBGs is
in fact highly significant (see Section~\ref{sec:models:dust}).  The
shortfall in LBG numbers that results when realistic dust extinction
is included (which we will quantify) motivates us to explore other
approaches, in particular, other modes of star formation.  We will
show how simply changing this ingredient can dramatically change our
predictions about the high redshift Universe, and discuss which --- if
any --- of the recipes can be ruled out by comparison with the
observational data.

We will define ``quiescent'' star formation to be the standard mode of
star formation that occurs within galactic discs whenever cold gas is
present. It is the dominant mode of star formation in the present-day
Universe, and there is observational evidence that its efficiency in
nearby galaxies is related to the internal properties of galactic
discs, apparently either to the surface density or to the dynamical
time \cite{kennicutt:98}.  Since both of these properties are slowly
varying, the star formation rate under quiescent star formation
changes only slowly with time.

We consider also a second mode of star formation in which stars are
created with dramatically increased efficiency over relatively short
time scales, termed ``starbursts.''  Various physical mechanisms might
produce such bursts, but here we will assume that they are triggered
by galaxy-galaxy mergers. We term these ``collisional'' starbursts to
distinguish them from bursts triggered by other means.  Substantial
observational and theoretical evidence exists in support of the
collisional starburst phenomenon. For example, a strong correlation is
observed between starburst activity and interactions in local galaxies
\cite{kennicutt:98}, and high-resolution N-body simulations including gas
dynamics
\cite{mihos:94,mihos:95,mihos:96,barnes:96} have shown that collisions
can trigger a bar-like instability that efficiently drives gas into
centrally concentrated, high density knots, creating the conditions
that are likely to result in a starburst. 

Our terminology implies a dichotomy between ``quiescent'' and
``bursting'' star formation that may be somewhat artificial ---
the recent work of
\citeN{kennicutt:98} suggests that star formation in 
both quiescent (normal) and starburst galaxies has the same dependence
on the \emph{local} gas density. By funneling gas into a small central
region, the interaction may simply produce the elevated gas densities
that in turn lead to enhanced levels of star formation. Still, in
interpreting the high redshift observations it is important
conceptually to determine what kind of process actually dominates
early star formation; the difference between internally-governed
versus externally-triggered star formation is highly significant.  For
example, it determines whether we interpret the number and luminosity
density of high redshift galaxies as reflecting basically their masses
and internal properties, or instead the merger rate at that epoch and
the efficiency of the gas inflows produced by these mergers.

In this paper, we investigate a scenario in which most of the
observable LBGs are collisional starbursts.  Despite the fact that
such collisions inevitably arise in hierarchical clustering models
(e.g., \citeNP{kolatt:00}), small-object collisions would not have
been resolved in existing hydrodynamical simulations
(for example \citeNP{katz:99}), while previous semi-analytic work may have
also underestimated the importance of collisional starbursts because
of their assumed physical recipes.  We investigate whether the numbers
and properties of high-redshift starbursts are compatible with
observations of LBGs, and whether this scenario leads to
self-consistent agreement with global quantities such as the star
formation rate density and the cold gas and metal content of the
Universe as a function of redshift. We also consider models containing
only ``quiescent'' star formation and discuss whether the collisional
starburst picture is merely
\emph{compatible} with the data or whether some contribution from
bursts seems to be
\emph{required}.
Our conclusion is that quiescent models can marginally match all data
provided that early star formation is highly accelerated, but that
burst models fit all data naturally and are therefore preferred.
Either way, we conclude that early star formation must be much more
efficient than locally.

The outline of this paper is as follows. In Section~\ref{sec:models},
we give a brief introduction to the semi-analytic models, including
our simple approach for including collisional starbursts. In
Section~\ref{sec:results}, we present the predictions of our fiducial
models and compare them with the observations.  In
Section~\ref{sec:previous}, we compare our results with previous work,
and in Section~\ref{sec:variations} we discuss the sensitivity of our
results to various assumptions, including the cosmology, stellar
population synthesis and dust modelling. We summarize our results and
conclude in Section~\ref{sec:conclusions}. The non-specialist reader
may wish to skip Sections~\ref{sec:previous} and
\ref{sec:variations}, which are somewhat technical. The main results
of these sections are summarized in a general way in
Section~\ref{sec:conclusions}.

\section{Semi-Analytic Modelling}
\label{sec:models}
\subsection{Basics}
\label{sec:models:basic}
We use semi-analytic techniques to model the formation and evolution
of galaxies in a hierarchical clustering framework. Our models include the
effects of gravitation on the formation and merging of dark matter
haloes, the hydrodynamics of cooling, star formation, supernovae
feedback and metal production, galaxy-galaxy merging, and the
evolution of stellar populations.  Our Monte Carlo-based approach is
similar in spirit to the models originally presented by
\citeN[hereafter KWG93]{kwg:93} and
\citeN[hereafter CAFNZ94]{cafnz:94}, and subsequently developed by
these groups (hereafter referred to as the ``Munich'' and ``Durham''
groups) and others in numerous papers. The semi-analytic models
used here are described in
\citeN{mythesis} and \citeN[hereafter SP]{sp}. As shown in SP,
these models produce good agreement with a broad range of local galaxy
observations, including the Tully-Fisher relation, the B and K-band
luminosity functions, cold gas contents, metallicities, and colours. We
refer the reader to SP for a more comprehensive review of the
literature and a more detailed description of our models. Below we
sketch our approach briefly.

The framework of the semi-analytic approach is the ``merging history''
of a dark matter halo of a given mass, identified at $z=0$ or any
other redshift of interest. We construct Monte-Carlo realizations of
``merger trees'' using the method of \citeN{sk}, which was tested
against merger trees extracted from dissipationless simulations
\cite{slkd}. 
Each branch in the tree represents a halo merging event, and the trees
are truncated when the circular velocity of the progenitor halo
becomes smaller than 40 \kms. We assume that gas in halos smaller than
this effective mass resolution is photoionized and cannot cool or form
stars (see SP). We construct merger histories for a grid of halos with
circular velocities ranging from 40 \kms to 1500
\kms, and weight the results using the appropriate
Press-Schechter probability for the appropriate halo mass and
redshift.

When a halo collapses or merges with a larger halo, we assume that the
associated gas is shock heated to the virial temperature of the
halo. This gas then radiates energy and cools. The cooling rate
depends on the density, metallicity, and temperature of the gas. Cold
gas is turned into stars using a simple recipe, and supernovae energy
reheats the cold gas according to another recipe. To model the star
formation rate, we use an expression of the general form
\begin{equation}
\label{eqn:sfr}
\dot{m}_{*} = \frac{m_{\rm cold}}{\tau_{*}}\,,
\end{equation}
where $\dot{m}_{*}$ is the star formation rate, $m_{\rm cold}$ is the
total mass of cold gas in the disc, and cold gas is converted into
stars with a time scale $\tau_{*}$. In principle this time scale could
be a function of the circular velocity or dynamical time of the disc
or other variables. In the simplest version of this recipe, $\tau_{*}
=\tau^0_{*} =$ constant; i.e. the efficiency of conversion of cold gas
into stars $\dot{m}_{*}/m_{\rm cold}$ is independent of the other
properties of the galaxy.  This recipe was one of those investigated
in SP (called SFR-C in the terminology of SP; here called ``constant
efficiency quiescent''). We also consider a recipe in which $\tau_{*}
\propto t_{\rm dyn}$, where $t_{\rm dyn}$ is the dynamical time of the
galaxy. Because the density of the collapsed haloes is higher at high
redshift, the typical dynamical times become smaller and the
conversion of gas into stars becomes faster. This recipe is similar to
the one usually used by the Munich group (called SFR-M in SP), and
here we call it ``accelerated quiescent.''

Supernovae feedback is modelled using the disc-halo model introduced
in SP. The rate of reheating of cold gas is given by
\begin{equation}
\label{eqn:fb}
\dot{m}_{\rm rh} = \frac{\dot{\epsilon}_{\rm SN}}{v^2_{\rm esc}}, 
\end{equation}
where $\dot{\epsilon}_{\rm SN}$ is the rate at which energy is injected
into the cold gas by supernovae, and $v^2_{\rm esc}$ is the average
escape velocity of the disc or halo (the reheating rate and ejected
gas mass is calculated seperately for each of these components).

Each new generation of stars produces a fixed yield of metals, which
are immediately deposited back into in the cold gas. The metals may
then be mixed with the hot gas in the halo, or ejected from the halo
by supernovae feedback. This in turn affects the cooling, which
becomes more efficient as the metallicity of the gas increases. Unlike
our work in SP, where we used a fixed metallicity for the hot gas in
calculating the cooling function, here we use the modelled value of
the hot gas metallicity and the metallicity-dependent radiative
cooling curves tabulated by
\citeN{sutherland:93}.

Another new feature (not present in the models of SP), is the
treatment of gas and metals that are ejected from the halo. In SP this
material was never re-incorporated in any halo. Here, we assume that
the material is distributed outside of the halo with a continuation of
the isothermal $r^{-2}$ profile that we assume inside the halo, and
such that if the total mass of the halo were to double, all of the
material would be re-incorporated in the halo. This material falls in
gradually (according to a spherical infall model) as the virial radius
of the halo increases due to the falling background density of the
Universe.  This is similar to the recipe used by CAFNZ and
\citeN{kauffmann:99}, in which all of the hot gas ejected from the
halo is re-incorporated (all at once) when the halo doubles in mass,
but it removes the discontinuous behaviour caused by the sudden infall
of a large amount of gas. The assumption that the ejected gas should
return when the halo mass doubles is just an arbitrary relic of the
block model used by CAFNZ, in which the halo masses always grew by
factors of two in each branch of the merging tree; the actual
timescale for return of the ejected gas is quite uncertain.

We now also account for the mass that is returned to the ISM by young
stars. For each quantity of mass that is turned into stars, a fraction
$R$ is returned to the cold gas reservoir. Here we assume $R=0.14$,
which results from assuming that all mass from stars with masses
greater than 8 \msun\ is recycled, using a Salpeter IMF with an upper
mass cutoff of 100 \msun. Had we also included mass loss from
smaller-mass stars, the value of $R$ would be larger, but these
estimates are more uncertain so we neglect this contribution.

\subsection{Mergers and Starbursts}
\label{sec:models:bursts}
When dark matter haloes merge, the galaxies contained within them
survive and may merge on a longer time scale. After a halo merger
event, the central galaxy of the largest progenitor halo becomes the
new central galaxy, and all other galaxies become
``satellites.'' Satellite galaxies may merge with the central galaxy
on a dynamical friction time scale, or with each other on
approximately a mean free path time scale. However, if the relative
velocities of the satellites are large compared to their internal
velocities, they will not experience a binding merger. Thus the
satellite merger rate decreases in clusters. The expressions for the
satellite-central and satellite-satellite merger rates are given in
SP.

We assume that every galaxy-galaxy merger triggers a starburst. 
Our treatment is based on a simple parameterization of the
results of N-body simulations with gas dynamics, which we now
summarize briefly. Mihos \& Hernquist \citeyear{mihos:94,mihos:96}
have simulated galaxy-galaxy mergers using a high resolution
N-body/smoothed particle hydrodynamics code (TREESPH) with star
formation modelled according to a Schmidt law ($\rho_{SFR} \propto
\rho_{\rm gas}^{n}$ with $n=1.5$). In \citeN{mihos:96}, mergers
between equal-mass galaxies (major mergers) were simulated, and it was
found that 65-85 percent of the total gas supply (in both galaxies) was
converted into stars over a time scale of 50-150 Myr. These results
were not very sensitive to morphology or the orbital geometry.

The case of highly unequal-mass mergers was explored in
\citeN{mihos:94}.  The case simulated represents a Milky Way-sized 
disc galaxy accreting a satellite that is one tenth of its
mass. The non-axisymmetric mode generated by the accretion of the
satellite causes a large fraction of the gas to collapse into the
central region of the galaxy, fueling a strong starburst. In this case
about 50 percent of the original gas supply is consumed in the
starburst, which lasts for about 60 Myr. However, the results are much
more sensitive to the morphological structure of the galaxies than the
major-merger case. If the larger galaxy has a bulge (the case
simulated has a bulge to disc mass ratio of 1:3), the bulge seems to
stabilize the disc against strong radial gas flows, leading to a much
weaker starburst event (only about 5 percent of the total gas supply
is consumed). \citeN{mihos:94} note that this implies that the
importance of bursts will decrease at low redshift as galaxies
develop bulges, even if the merger rate remains constant.

\begin{figure}
\centerline{\psfig{file=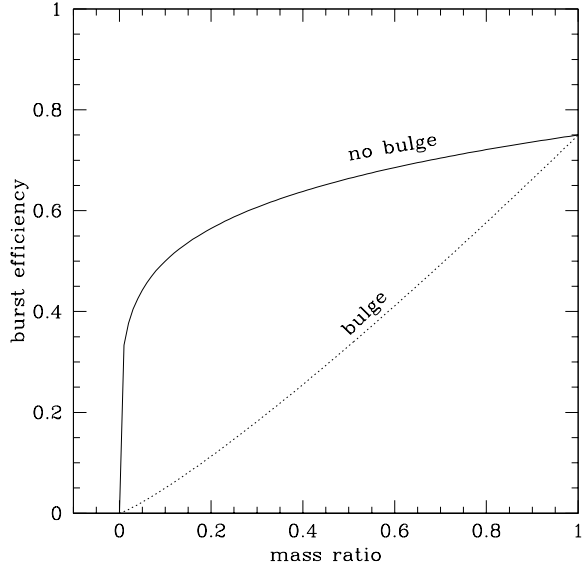,height=8truecm,width=8truecm}}
\caption{The efficiency of star formation in bursts as a function of
the baryonic mass ratio of the merging galaxies. The results are
scaled to match the calculations of \protect\citeN{mihos:94} and
\protect\citeN{mihos:96}. When a bulge is present, the burst efficiency
drops off more quickly at lower mass ratios. }
\label{fig:eburst}
\end{figure}

These results inform our treatment of collisional starbursts.  When
any two galaxies merge, the ``burst'' mode of star formation is turned
on. The star formation rate due to the burst is modelled as a Gaussian
function of time 
with a width of $\sigma_{\rm burst}$. The burst is
switched off after a time $4 \sigma_{\rm burst}$ has elapsed. The
burst model has two parameters, the time scale and the efficiency of
the burst. The efficiency $e_{\rm burst}$ is defined as the fraction
of the cold gas reservoir (of both galaxies combined) that is turned
into stars over the entire duration of the burst. We assume that the
efficiency is a power-law function of the mass-ratio of the merging
galaxy pair:
\begin{equation}
e_{\rm burst} = f_{\rm consume} \left[m_{\rm small}/m_{\rm
big}\right]^{\alpha_{\rm burst}}\, ,
\end{equation}
with the two parameters $f_{\rm consume}$ and $\alpha_{\rm burst}$
chosen to reproduce the results of
\citeN{mihos:96} for a 1:1 merger, and \citeN{mihos:94} for a 1:10
merger ($f_{\rm consume}=0.75$ and $\alpha_{\rm burst}=0.18$). When a
bulge of at least a third of the disc mass is present, bursts are
suppressed in minor mergers (based on \citeNP{mihos:94}).
Fig.\ref{fig:eburst} shows the resulting scaling of the burst
efficiency with the mass ratio of the merging pair for the bulge and
no-bulge cases. We assume that the burst timescale $\sigma_{\rm
burst}$ is proportional to the dynamical time of the larger of the two
discs. The quiescent mode of star formation continues uninterrupted
according to Eqn.~\ref{eqn:sfr} above, although its contribution is
generally insignificant in comparison to the burst.

Each merger is classified as ``major'' or ``minor'' according to
whether the ratio of the smaller to the larger of the galaxies'
baryonic masses is greater than or less than the value of the
parameter $f_{\rm bulge} \sim 0.25$.  Major mergers have mass ratios
greater than $f_{\rm bulge}$, and the bulge and disc stars of both
galaxies plus all new stars formed in the burst are placed in a bulge
component. Minor mergers have mass ratios less than $f_{\rm bulge}$,
and the pre-existing stars from the smaller galaxy are placed in the
disc component of the post-merger galaxy while all newly formed stars
again join the bulge.  This is motivated by the N-body simulations of
similar satellite accretion events by
\citeN{walker:96}, which show that 90 percent of the satellite mass is
stripped and distributed in the disc of the larger galaxy and only
the densest part of the satellite core ends up at the centre. 

A possible problem in scaling from the $z=0$ simulations is that
galactic discs may be quite different at high redshift. Galaxies then
are smaller and denser and probably have higher ratios of gas to
stars. This may significantly affect their susceptibility to
star-forming instabilities.  We are currently investigating this using
a more extensive set of N-body hydrodynamic simulations similar to
those of Mihos \& Hernquist, but with the initial galaxy properties
chosen to be representative of high-redshift galaxies
\cite{somerville:00}. Similarly, the modelling of mergers,
infall and tidal stripping of sub-haloes should be refined and tested
against high-resolution dissipationless cosmological simulations. Work
on these issues is also in progress \cite{kolatt:00}.

\begin{figure}
\centerline{\psfig{file=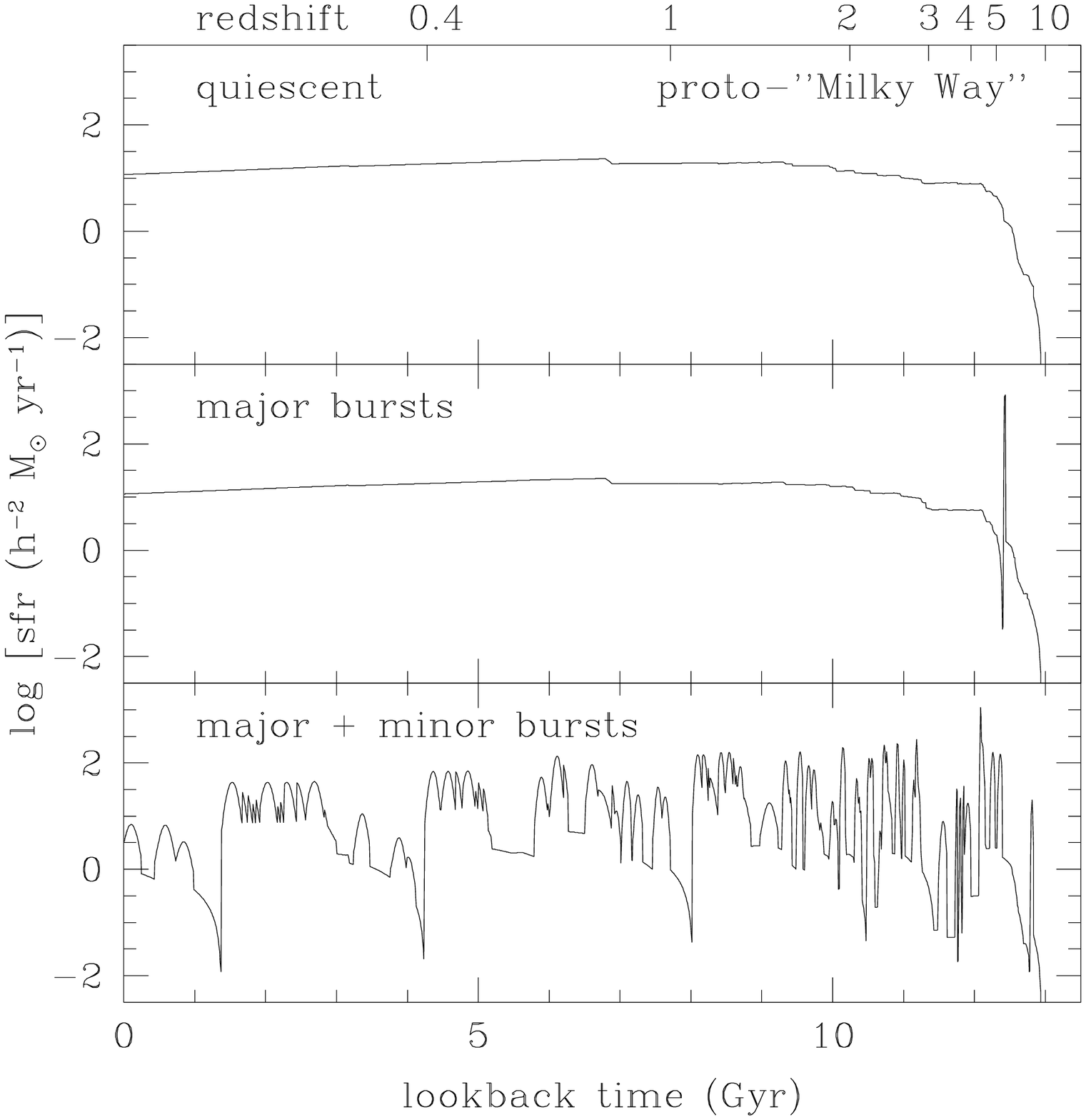,height=10truecm,width=8truecm}}
\centerline{\psfig{file=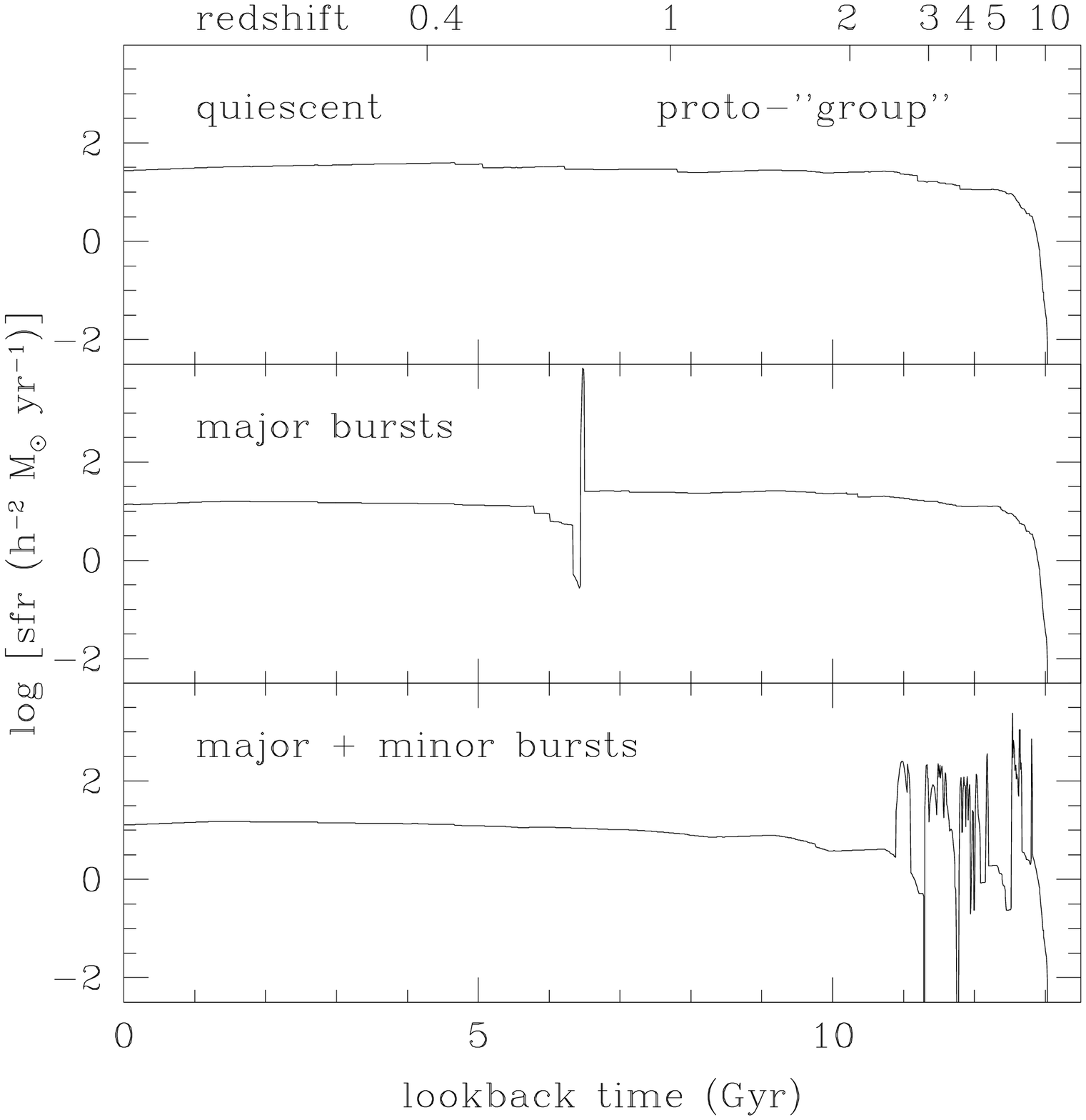,height=10truecm,width=8truecm}}
\caption{The total star formation rate for the largest progenitor of
the central galaxy within a halo with a present-day circular velocity
of 220 or $500\, \kms$.  All three models contain constant efficiency
quiescent star formation. The top panel shows the model with no
bursts, the middle panel shows the model with bursts in major mergers
only, and the bottom panel shows the models with bursts in major and
minor mergers.}
\label{fig:sfhist}
\end{figure}

Fig.~\ref{fig:sfhist} shows the total star formation rate for the
largest progenitor of the central galaxy in a Milky Way-sized halo
($V_c = 220\,\kms$) and in a group-sized halo ($V_c = 500\,\kms$),
both at $z=0$ today.  The star formation rate is shown in models with:
(1) no starbursts, (2) bursts in major mergers only, and (3) bursts in
both major and minor mergers. All three models contain quiescent star
formation using the constant efficiency recipe.  From this figure we
can see that minor mergers are important even if their efficiency is
low because they are much more common than major mergers.
Comparing the group-sized halo with the Milky Way halo shows how
mergers shut down in high-velocity-dispersion environments.  This is
because we have assumed that the relative velocity of colliding
objects must be small compared to their internal velocities in order
for objects to merge.  The burst models discussed in the remainder of
this paper correspond to model (3) unless specified otherwise.

\subsection{Stellar Population Synthesis}
With the full star formation history of a galaxy in place, we estimate
its luminosity in any desired passband using the stellar population
synthesis models of Bruzual \& Charlot (GISSEL). In this paper, our
standard treatment uses the 1998 version of the solar-metallicity
models with a Salpeter IMF.  Modelling the young, massive, sub-solar
metallicity stars that might be found in LBGs is probably quite
uncertain (see
\citeNP{cwb}). However, it is encouraging that several 
groups have produced models that agree quite well in the UV and
optical part of the SED (see \citeNP{devriendt} and
Section~\ref{sec:variations:sed}). The mass-to-light ratio in the UV
is relatively insensitive to metallicity for young stars, but it is
quite sensitive to the IMF (primarily to the ratio of high-mass to
low-mass stars). We include only starlight and neglect the
contribution from nebular emission.  See
Section~\ref{sec:variations:sed} for a detailed discussion of
uncertainties in the stellar population synthesis.

\subsection{Dust}
\label{sec:models:dust}
Because most of the observations of high-redshift galaxies are
obtained in the rest-UV, the effects of dust are likely to be
important. Just \emph{how} important has been a matter of debate ever
since the discovery of the LBG population. \citeN{pettini:97}
estimated an extinction at $\sim1500$ \AA\ of a factor of $\sim 3$,
based on the ratio of emission lines to the continuum in a few
objects. \citeN{meurer:97} and
\citeN{sawicki:98} estimated much larger factors of 15-20. Recently,
there seems to have been convergence to an intermediate value of a
factor of $\sim 5$ \cite{meurer:99,steidel:99}. This work makes use of a
correlation between the FIR excess (a reliable observational
measure of bolometric extinction) and the far-UV spectral slope in nearby
starburst galaxies. If the same correlation is then assumed to hold in
high-redshift galaxies, the measured UV spectral slopes of LBGs
indicate that there is an average extinction of a factor of 4.7 in the
relatively bright (${\mathcal R} < 25.5$) LBG population studied by
\citeN{steidel:99}. The same measurements \cite{meurer:99} 
indicate that, in LBGs as in local starburst galaxies, the most
UV-luminous (and rapidly star forming) objects are the most heavily
extinguished \cite{wh}. 

To estimate the effects of dust on our model galaxies, we use a very
simple parameterization of the empirical results of
\citeN{wh} for nearby starburst galaxies. The optical depth for a
galaxy with an intrinsic (unextinguished) UV luminosity $L_{UV, i}$ is
given by:
\begin{equation}
\tau_{UV} = \tau_{UV,*}\left(\frac{L_{UV, i}}{L_{UV, *}}\right)^\beta\, .
\end{equation}
This is identical to the recipe used in SP except that we now
normalize the recipe in the UV rather than the B band. We take $L_{UV,
*}$ to equal the observed value of $L_{*}$ in the $z=3$ sample
of
\citeN{steidel:99}, i.e., the luminosity corresponding to
$m_{AB}=24.48$ in the appropriate cosmology. We use the same value of
$L_{*}$ at redshift $z=4$, which as \citeN{steidel:99} note is
consistent with the $z=4$ data. We take $\beta=0.5$ as in
\citeN{wh}. We then assign to each galaxy a random inclination and
compute the actual extinction using a standard slab model (see SP). We
adjust the parameter controlling the face-on optical depth,
$\tau_{UV,*}$, to obtain an average extinction correction at 1500 \AA\
of a factor of $\sim 5$ for $z\sim3$ galaxies with luminosities
typical of the Steidel et al. sample, consistent with the estimates
described above. The value $\tau_{UV,*}=1.75$ gives good results. To
extend the extinction correction to other wavebands, we assume a
Calzetti attenuation curve \cite{calzetti:97a}.

\subsection{Model Parameters}
Throughout this paper, unless stated otherwise, we use the fashionable
\lcdm\ cosmology with $\Omega_0=0.3$, $\Omega_{\rm\Lambda}=0.7$,
$h\equiv H_0/(100 {\rm km/s/Mpc}) = 0.7$, $\sigma_8=1.0$.  These
values are consistent with a great deal of current data (for a recent
review see \citeNP{primack:00}). As in SP, the main free parameters
related to galaxy formation --- those describing the quiescent star
formation efficiency ($\tau^{0}_{*}$), the supernovae feedback
efficiency ($\epsilon^{0}_{\rm SN}$), and the metallicity yield ($y$)
--- are set by requiring an average present-day ``reference galaxy''
(the central galaxy in a halo with a circular velocity of $220
\,\kms$) to have (a) an I-band magnitude $M_{I} -5\log h = -21.8$, (b)
a cold gas mass $m_{\rm cold}
\simeq 10^{10} h^{-2} \msun$, and (c) a stellar metallicity of
about solar. Requirement (a) adjusts the zero-point of the I-band
Tully-Fisher relation to agree with observations, while
requirement (b) fixes the cold gas content of the ``reference galaxy''
to agree with observed H$_{\rm I}$ masses, allowing for a
contribution from molecular hydrogen (see SP). The values of the free
parameters used here are similar to those used in SP.

In some previous work (e.g., BCFL), it has been assumed that only a
fraction $f^{*}_{\rm lum} < 1$ of the stars formed in the models are
luminous, the remainder being in the form of brown dwarfs or other
non-luminous baryonic material.  We find that in order to get our
reference galaxy to be bright enough today with our assumed Salpeter
IMF, we need $f^{*}_{\rm lum} \simeq 1$.

Note that the parameters have been set entirely by comparing to a
subset of present-epoch observations. We have shown in SP that the
models then reproduce many important observed features of nearby
galaxies. Now, leaving all free parameters fixed, we can consider
the predictions of the same models for the high-redshift Universe.

\section{Model Results}
\label{sec:results}
Because extensive spectroscopic follow-up work has now verified the
effectiveness of the Lyman-break or ``drop-out'' technique for finding
$z \ga 2$ galaxies, we do not attempt to mimic the colour selection
process used to identify real Lyman-break objects but rather assume
that all galaxies brighter than a certain limiting magnitude in our
models would be in fact be selected as LBGs. The modelling work of
BCFL has shown that using the same colour selection criteria as the
observations picks out model galaxies in the expected redshift range.
We do not include the effect of absorption by intervening cold gas
clouds because, although this effect is very important shortwards of
the Lyman limit, for redshifts less than $z\sim4$ it does not much
affect the spectral energy distribution (SED) in the wavelength range
relevant to our study. We have calculated magnitudes using the filter
response functions of the WFPC2 filters F606W and F814W ($V_{606}$ and
$I_{814}$), as well as the ${\mathcal R}$ and $I$ filters of
\citeN{steidel:92} used in the ground based observations and kindly 
provided to us in electronic form by K. Adelberger. All magnitudes are
given in the AB system \cite{oke:83}. The $V_{606}$ filter and the
${\mathcal R}$ filter correspond to a mean rest wavelength of 1600
\AA\ 
at redshifts 2.75 and 3.04 respectively. 
The $I_{814}$ and Steidel \& Hamilton I filters correspond to the same
rest wavelength at redshifts of about 4.00 and 4.13
respectively. 

We now consider
three different recipes for star formation. The model 
referred to as the ``collisional starburst'' model includes quiescent
star formation using the ``constant efficiency'' (CE) recipe, plus
star formation in bursts as described in the previous
section. The ``constant efficiency quiescent'' model has no
burst mode, and all star formation is modelled using the constant
efficiency recipe.  The third model is also quiescent,
but star formation efficiency varies inversely
as the dynamical time of the galaxy. We refer to this third
model as ``accelerated quiescent'' because the star formation rate for
a given cold gas mass becomes
\emph{higher} at high redshift due to the increasing density of the
haloes. Note that the supernovae feedback recipe has been kept the
same and all other free parameters have been left fixed to the same
values for the three models.  In SP, we showed that, by adjusting only
the normalization of the star formation efficiency, $\tau^0_{*}$, it
is possible to make all three models to match observed properties of
local galaxies such as the B and K band luminosity functions, the
Tully-Fisher relation, colours, gas and metal contents, etc.

\begin{figure}
\centerline{\psfig{file=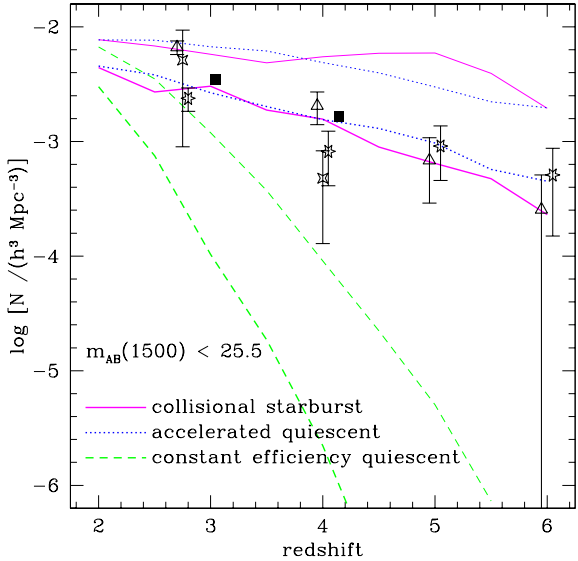,height=8truecm,width=8truecm}}
\caption{The comoving number density of galaxies brighter than 25.5 at 
rest 1500 \AA. Thin lines show the three models with no correction for
dust extinction, and bold lines show the models with dust extinction
included using the recipe described in
Section~\protect\ref{sec:models:dust}. The filled squares indicate the
comoving number density of LBGs with spectroscopic redshifts from the
ground-based sample of
\protect\citeN{steidel:99}. 
Four-pointed stars indicate the values derived from LBGs in the HDF
\protect\cite{pozzetti:98}. The stars and triangles are from the HDF-N
and HDF-S galaxies with photometric redshifts from the Stonybrook
catalogs \protect\cite{lanzetta:99a,chen:99a}. The observations have
not been corrected for extinction. Note that the constant efficiency
quiescent model predicts a strong decrease in the number density of
bright galaxies with redshift, in contrast to the other two models,
which show a very gentle decline from $z\sim2$ to $z\sim6$.}
\label{fig:counts}
\end{figure}
\subsection{Comoving Number Density}
We first investigate the model predictions for the number density of
bright objects as a function of redshift. Fig.~\ref{fig:counts} shows
the comoving number density of galaxies brighter than an apparent AB
magnitude limit of m$_{AB}(1500) =$ 25.5 at rest 1500 \AA \ over the
redshift range $2
\leq z \leq 6$. The comoving number densities for
the observations have been calculated
using the appropriate geometry for the \lcdm\
cosmology used in our models. 
For the ground-based sample of LBGs with
spectroscopic redshifts, we have used the values given in Table~3 of
\citeN{steidel:99}. The 1$\sigma$ field-to-field variance 
of this sample is about 12 percent; if plotted on
Fig.~\ref{fig:counts}, the error bar would be smaller than the
symbol. The comoving number density of LBGs in the HDF has been
calculated using the $V_{606}$ counts of U-dropouts from Table~1 of
\citeN{pozzetti:98}, and the $I_{814}$ counts of B-dropouts helpfully
provided to us by L. Pozzetti. Error-bars on these points represent
Poisson errors only. We also show the number densities computed from
the Stonybrook catalog of photometric redshifts from the HDF-N and
HDF-S \cite{lanzetta:99a,chen:99a}, generously provided to us by
H.-W. Chen and K. Lanzetta. Once again the error bars are simply
1$\sigma$ Poisson and do not account for photometric or redshift
errors. Note that, unlike the point derived from galaxies identified
in the HDF using the Lyman-break technique
\cite{pozzetti:98}, the values from the photometric redshift catalogs
at $z\sim3$ agree with the ground-based estimate within the
errors. This suggests that the full photometric redshift technique,
including the observed near-IR bands, finds high-redshift galaxies
that are missed by the Lyman-break technique. The difference between
the HDF-N and HDF-S (stars and triangles) gives an indication of the
field-to-field variance characteristic of the HDF volume.

Fig.~\ref{fig:counts} demonstrates an important and robust prediction
of the models --- the \emph{constant efficiency} quiescent models
predict a \emph{strong decline} in the number density of bright
galaxies with increasing redshift. This is a generic feature of any
hierarchial model in which the efficiency of the conversion of cold
gas into stars remains constant with redshift. This is because the
number density of the massive haloes that host bright galaxies in the
quiescent model drops off sharply with redshift. \citeN{kolatt:99}
found a similar result based on $N$-body simulations.  The net result
is that the CE quiescent model without dust extinction produces an
acceptable match to the number of bright galaxies at $z\sim 3$ but
badly underpredicts the numbers at redshifts $z\geq 4$. The shortfall
is unacceptable at all redshifts when dust is included.

In contrast, the burst model contains collisional starbursts that
cause smaller-mass objects to become bright enough to exceed the
detection limit for a brief amount of time. Because the collision rate
is relatively constant over the redshift range $2 \la z \la 6$ and
there is an ample supply of cold gas, the comoving number density of
visible galaxies remains \emph{almost constant} over this range.  The
collisional starburst model actually overpredicts the number of bright
galaxies, but adding dust provides excellent agreement at all
redshifts. Finally, the accelerated quiescent model produces nearly
identical results to the collisional starburst model, but for a very
different reason. Instead of being temporarily brightened by bursts,
each galaxy has a constant low mass-to-light ratio because of the
accelerated star formation rate.  This model also does a good job of
matching the number of moderately bright galaxies at all redshifts.

The observations at $2 \la z \la 4.5$, especially those based on the
large ground-based sample of \citeN{steidel:99}, provide a secure
lower limit on the number density of bright galaxies. The situation at
higher redshift is more uncertain, but the differences between the
constant efficiency quiescent model and the other two are quite
dramatic.  According to the CE quiescent model, the probability of
finding even one bright galaxy at $z\ga5$ in a volume comparable to
the HDF is only one in 10$^{6}$, whereas both the collisional
starburst model and the accelerated quiescent model predict on the
order of a few galaxies per HDF at $4.5
\la z \la 6$, even with dust. A number of candidate high-redshift galaxies
with $z \ga 5$ have been discovered
recently
\cite{dey:98,weymann:98,spinrad:98,chen:99b,vanbreugel:99,hu:99}, but,
as most of these objects were found serendipitously, it is difficult
to estimate their number density.  Such candidates are also being
detected in fair numbers in photometric redshift catalogs, but these
detections are also uncertain, and further spectroscopic work will be
required in order to confirm them (see \citeNP{stern:99} for a recent
review on search techniques for very high-redshift galaxies).  If the
current preliminary detections of very high-redshift galaxies are
confirmed, they will essentially rule out the constant efficiency
quiescent model.

\subsection{Luminosity Function}
\begin{figure}
\centerline{\psfig{file=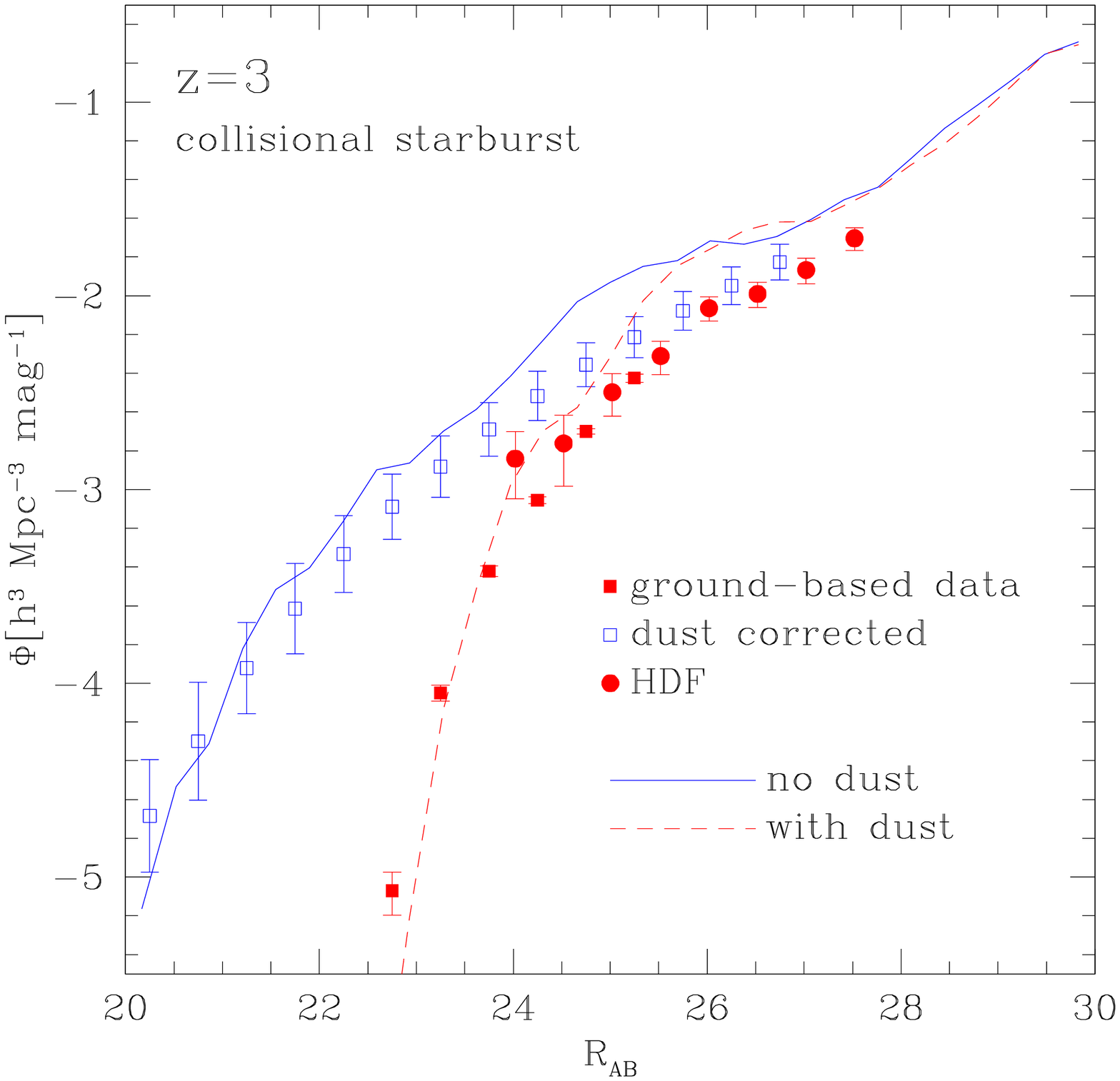,height=8truecm,width=8truecm}}
\caption{The luminosity function at $z=3$ in the observed ${\mathcal R}_{AB}$ band 
(rest $\sim 1600$ \AA). Symbols show the composite luminosity function
from ground-based observations (filled squares;
\protect\citeNP{steidel:99} and the HDF (filled circles;
\protect\citeNP{pozzetti:98}). Open squares
show the luminosity function derived from the ground-based
observations after correcting for the effects of dust extinction
using the method described in the text
\protect\cite{adelberger:00}. Lines indicate the results of the
collisional starburst model without (solid) and with (dashed)
corrections for dust extinction. }
\label{fig:lf_z3_burst}
\end{figure}
\begin{figure}
\centerline{\psfig{file=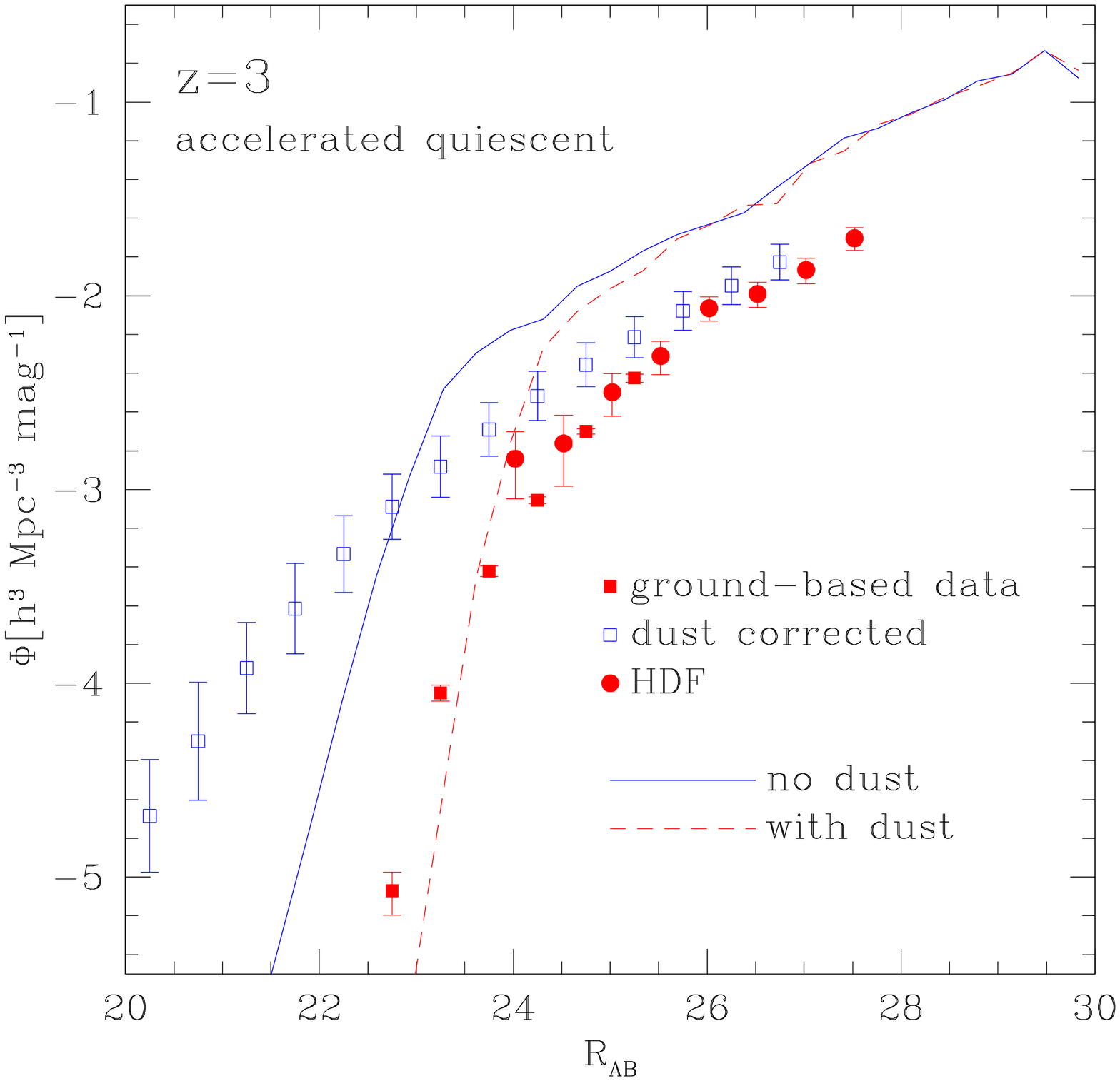,height=8truecm,width=8truecm}}
\centerline{\psfig{file=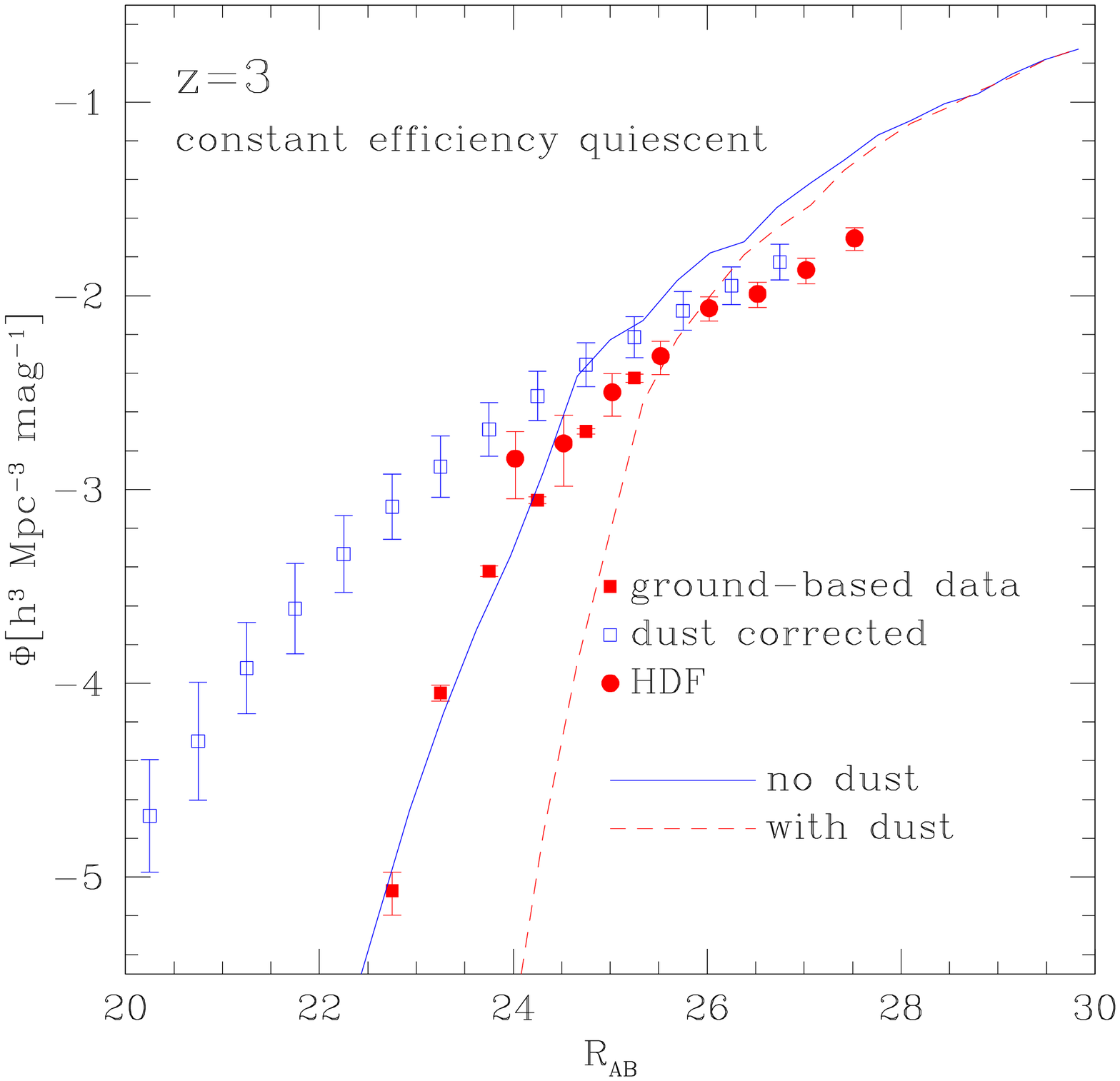,height=8truecm,width=8truecm}}
\caption{Same as Fig.~\protect\ref{fig:lf_z3_burst}, for the quiescent 
models. }
\label{fig:lf_z3_quiescent}
\end{figure}

\begin{figure}
\centerline{\psfig{file=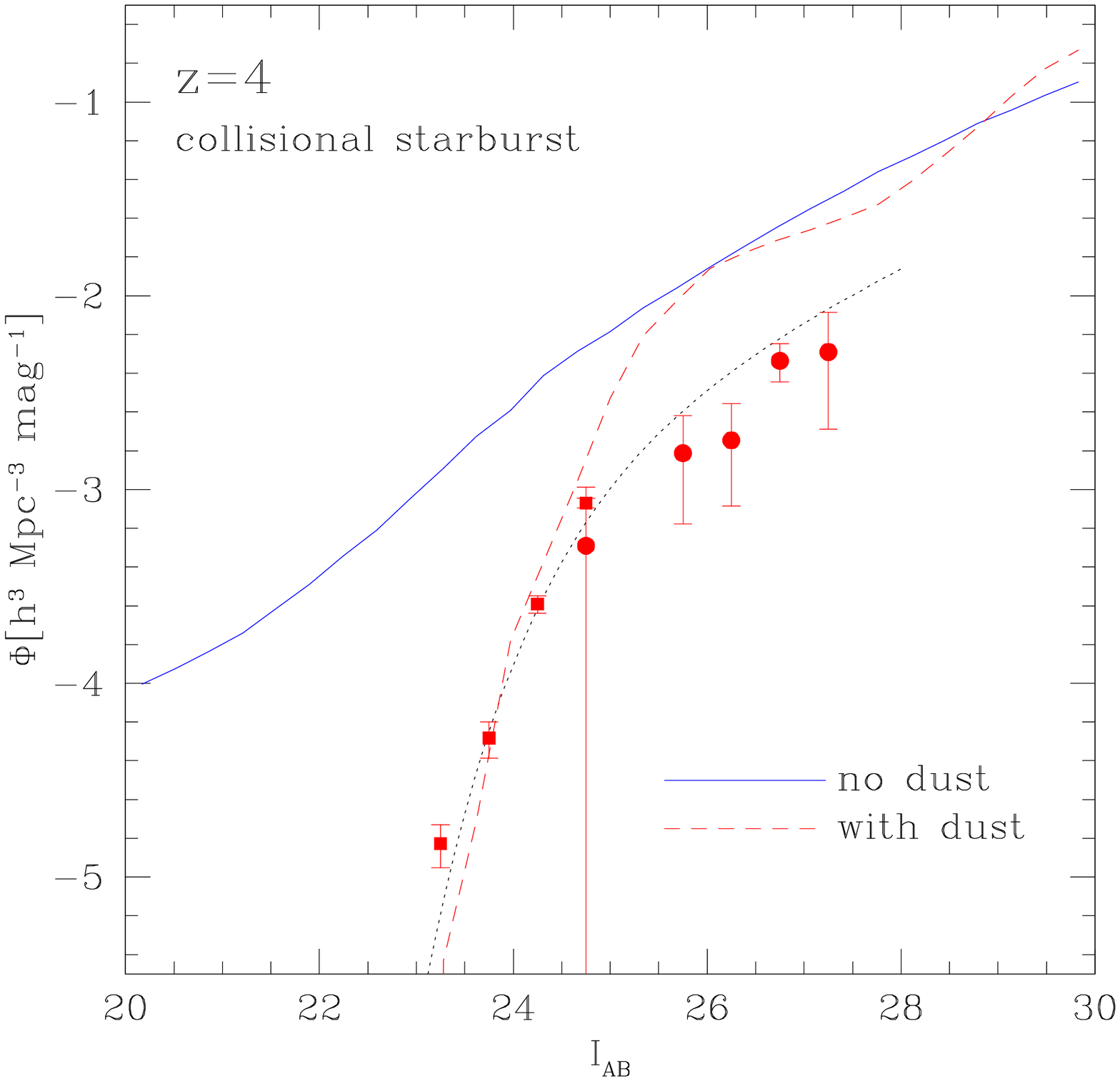,height=8truecm,width=8truecm}}
\caption{Same as Fig.~\protect\ref{fig:lf_z3_burst}, for the $I$ 
band at $z=4$ (rest $\sim 1600$ \AA).  Observed data are shown as solid
symbols (see key on Fig.~\ref{fig:lf_z3_burst}).  The dotted line
shows the Schechter fit to the $z=4$ luminosity function of
\protect\citeN{steidel:99}, which is somewhat higher than the HDF
results. 
The data may be affected by incompleteness below $\sim$26 magnitude.}
\label{fig:lf_z4_burst}
\end{figure}
\begin{figure}
\centerline{\psfig{file=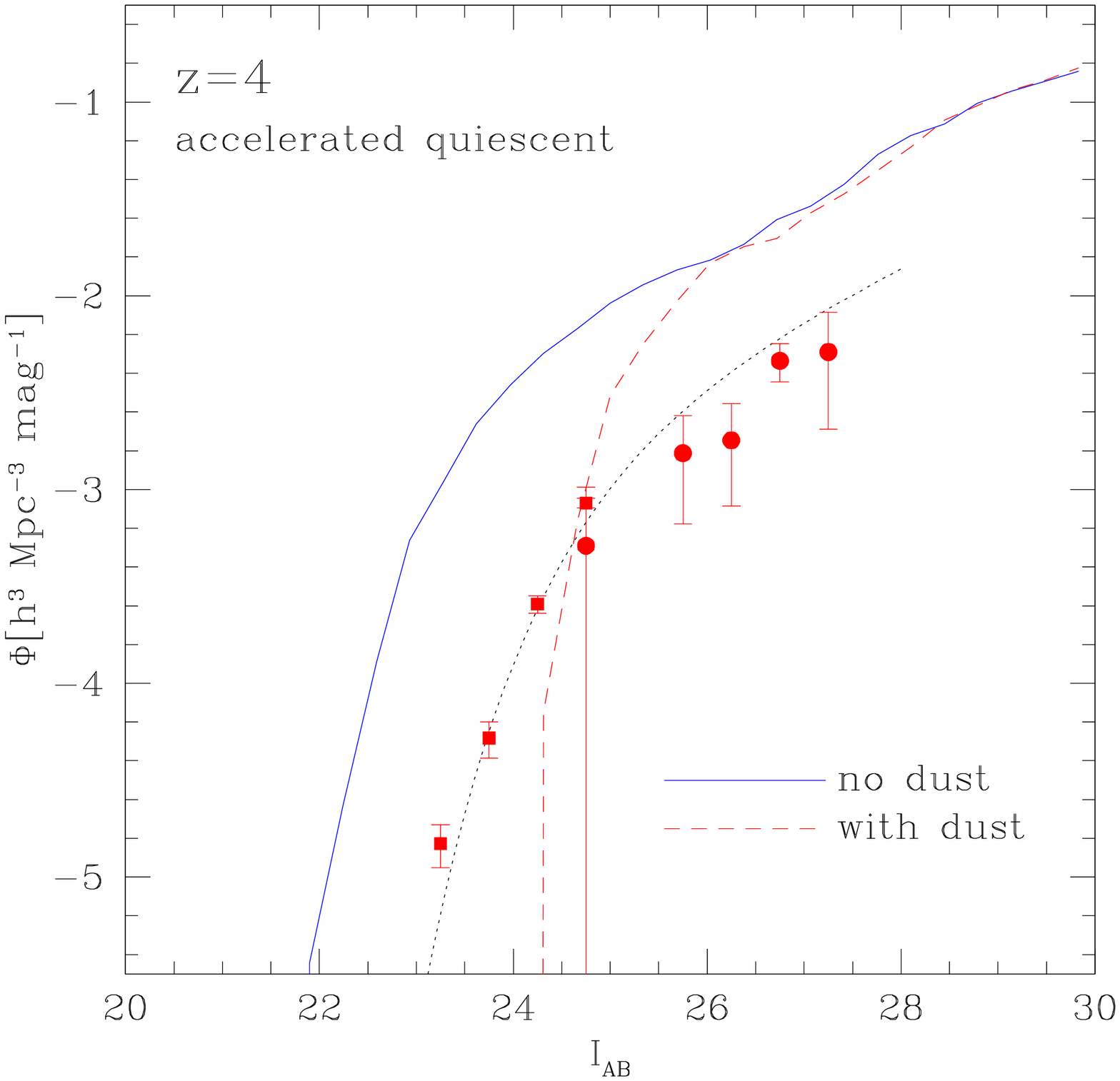,height=8truecm,width=8truecm}}
\centerline{\psfig{file=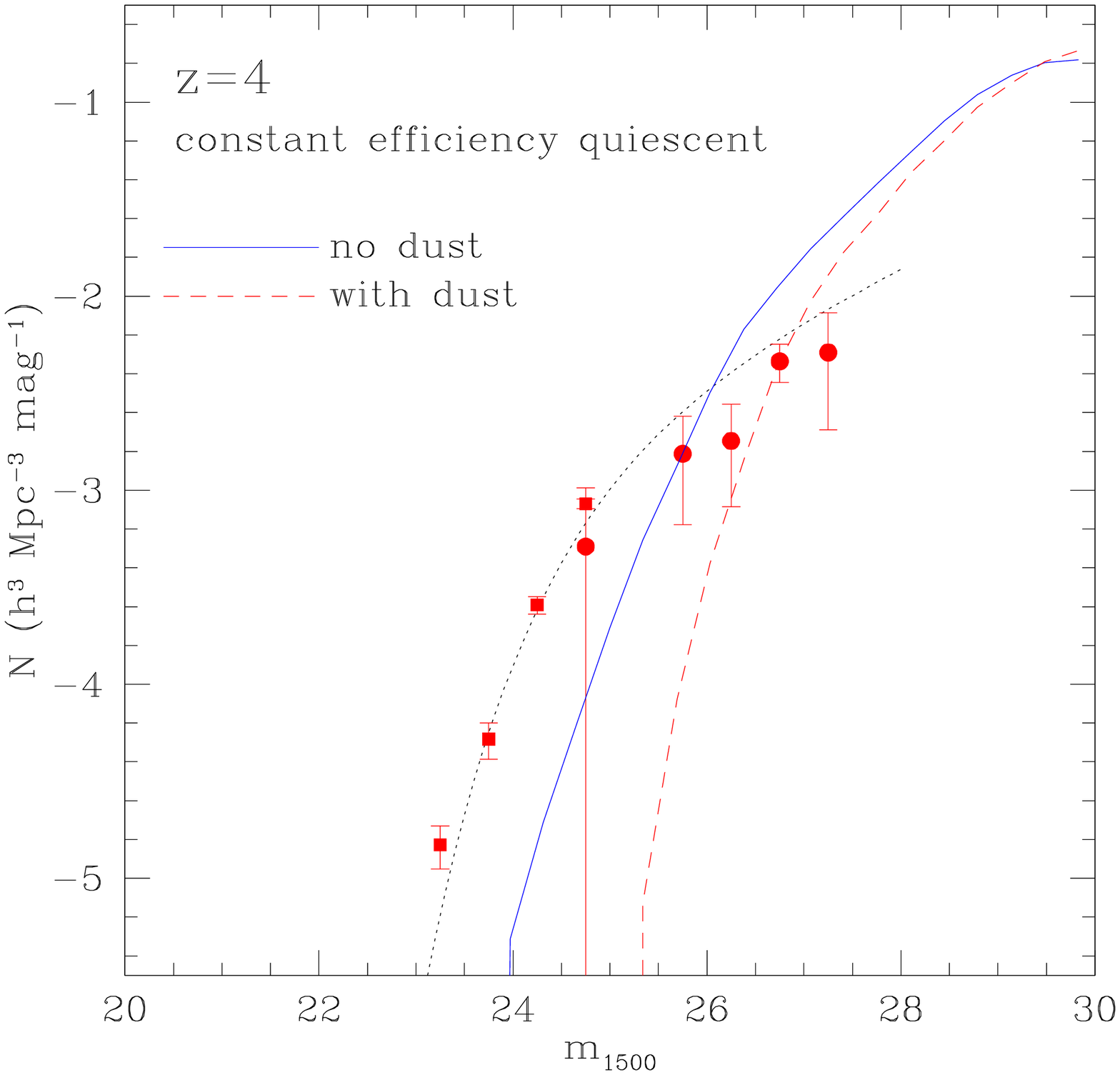,height=8truecm,width=8truecm}}
\caption{Same as Fig.~\protect\ref{fig:lf_z4_burst}, for the quiescent models. }
\label{fig:lf_z4_quiescent}
\end{figure}

The luminosity function and its evolution with redshift provide
additional tests of models.  The bright end ($m_{AB} \la 25.5$) of the
luminosity function at $\sim1500$ \AA\ for the observed LBGs at
$z\sim3$ can be determined fairly accurately from the relatively large
ground-based spectroscopic sample. The HDF can then be used to give
some indication of the faint end slope, although there is probably
some incompleteness faintwards of about $m = 26$
\cite{dickinson:98}. 
The resultant composite luminosity function at $z = 3$ is shown in
Figs. ~\ref{fig:lf_z3_burst} and ~\ref{fig:lf_z3_quiescent}.  The
ground-based data have been corrected for incompleteness (see
\citeNP{steidel:99}). As usual, the observations have been converted to our
\lcdm\ cosmology.

Fig.~\ref{fig:lf_z3_burst} shows that the luminosity function 
of the collisional starburst model with dust has a steep,
Schechter-type drop-off at bright magnitudes, whereas the intrinsic
function without dust is much flatter.
Thus in this picture, the knee seen in the observations is purely an
artifact of differential dust extinction.  It is gratifying that our
intrinsic luminosity function is very similar in shape to the observed
dust-corrected function derived from the data
\cite{adelberger:00}, for which the corrected luminosity of each
galaxy has been derived using the measured far-UV slope
and the dust-slope correlation discussed in \citeN{meurer:99}. 

Fig.~\ref{fig:lf_z3_quiescent} shows similar plots for the two
quiescent models.  The constant efficiency model without dust (bottom
panel) has a steeply decreasing luminosity function at the bright end,
reflecting the steepness of the halo mass function at these large
masses. When dust extinction is added, the CE quiescent model becomes
even steeper and dramatically underpredicts the number of bright
galaxies, consistent with its underprediction of count data in
Fig.~\ref{fig:counts}.  The situation with the accelerated quiescent
model is rather different.  Although that model predicts about the
same total number of bright ($m_{AB}(1500) < 25.5$) galaxies as the
collisional starburst model in Fig.~\ref{fig:counts}, the \emph{shape}
of its luminosity function is different at the very brightest
magnitudes. The burst model shows a tail at bright magnitudes, whereas
the accelerated quiescent model even \emph{without} dust drops off
steeply, like a Schechter function.  With dust included, the
accelerated quiescent model falls even more steeply and fails to match
the brightest galaxies.  This, again, is due to the tight link between
halo masses and galaxy luminosities in all quiescent models.

Figs.~\ref{fig:lf_z4_burst} and \ref{fig:lf_z4_quiescent} show
analogous comparisons for luminosity functions at $z \sim 4$.  The
collisional starburst model fits the luminosity function for bright
galaxies whereas both quiescent models fall short, especially when
dust is added.  All models predict too many faint galaxies.  This may
be due to the fact that, with the parameters we have chosen, our dust
model predicts very little extinction in galaxies fainter than $m
\simeq 26$.  In correcting the models for dust, we have simply assumed
the same parameters as at $z\sim3$.  Note also that the observed
luminosity function faintwards of $m
\simeq 25$ (the magnitude reached by the ground-based survey) is quite
uncertain and possibly incomplete.  Both effects would tend to make
the models deviate in this way.

In summary, the CE quiescent models predict a strong evolution of
$L_{*}$ with redshift, wheras the burst model and the accelerated
quiescent model predict very little evolution in $L_{*}$ over the
redshift range $3
\la z \la 5$.
The observed and dust-corrected luminosity functions of the
collisional starburst model are in good agreement with all data at
both redshifts.  However, systematic errors in the dust correction
could be large.  A more direct test of the number densities of very
bright galaxies (which correspond to objects with extremely high star
formation rates) will come from future sub-millimeter observations.

\subsection{The Star Formation History of the Universe}
The star formation history of the Universe, i.e., the global rate of
star formation as a function of redshift, is a crucial test of
theories of galaxy formation and cosmology. There are many
observational tracers of the star formation history, each with their
own selection effects. In this section we investigate the star
formation history traced by the luminosity density, age constraints on
stars in nearby galaxies, cold gas detected in quasar absorption
systems, and metals.

\subsubsection{Luminosity Density}
\label{sec:results:sfrd}
\begin{figure}
\centerline{\psfig{file=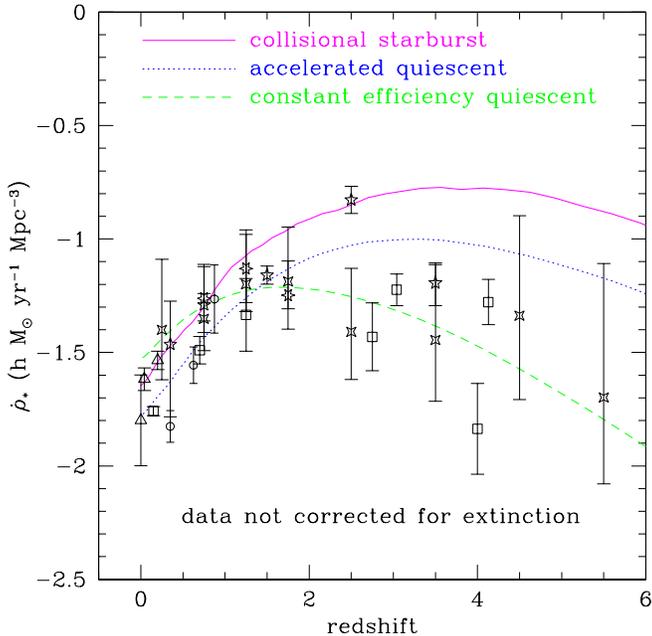,height=9truecm,width=9truecm}}
\caption{The star formation history of the Universe 
(``Madau diagram'') for our \lcdm\ cosmology.  Symbols indicate
observed estimates of the star formation rate density at various
redshifts; sources of data are listed in Table A2.  All data points
have been corrected for incompleteness but \emph{not} for dust
extinction.  Lines show predictions of the three models. }
\label{fig:new_madau_nodust}
\end{figure}
\begin{figure}
\centerline{\psfig{file=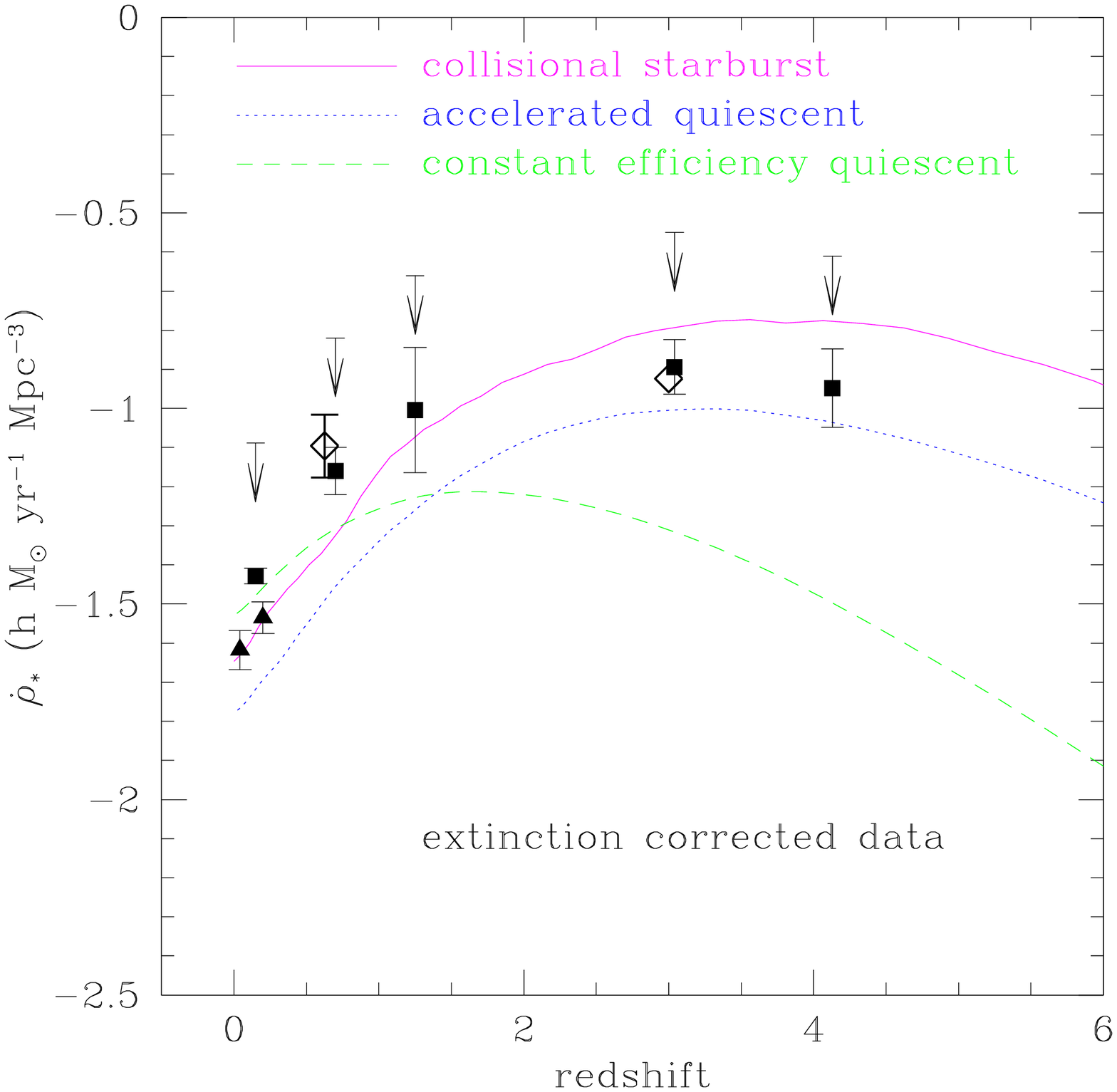,height=9truecm,width=9truecm}}
\caption{The new Madau diagram, corrected for dust extinction.
Symbols indicate observational estimates of the star formation rate
density at various redshifts. Triangles are derived from H$\alpha$
\protect\cite{kiss,tresse:98},
and squares are derived from rest-UV luminosities
\protect\cite{treyer:98,cowie:99,steidel:99}. These data have been 
corrected for incompleteness and dust extinction (see Appendix for
details).  The upper-limit symbols show the data with a ``maximal dust
correction,'' in which a fixed factor of five correction has been
applied to all galaxies.  If dust extinction is luminosity dependent,
this probably overestimates the true extinction correction (see
text). Diamonds indicate results derived from far-IR sources at
$z=0.7$ observed by ISO
\protect\cite{flores:99} and sub-mm sources,
believed to be at high redshift ($\sim 3$), observed by SCUBA
\protect\cite{hughes:98}. Lines show the three usual models. }
\label{fig:newmadau_dust}
\end{figure}
\begin{figure}
\centerline{\psfig{file=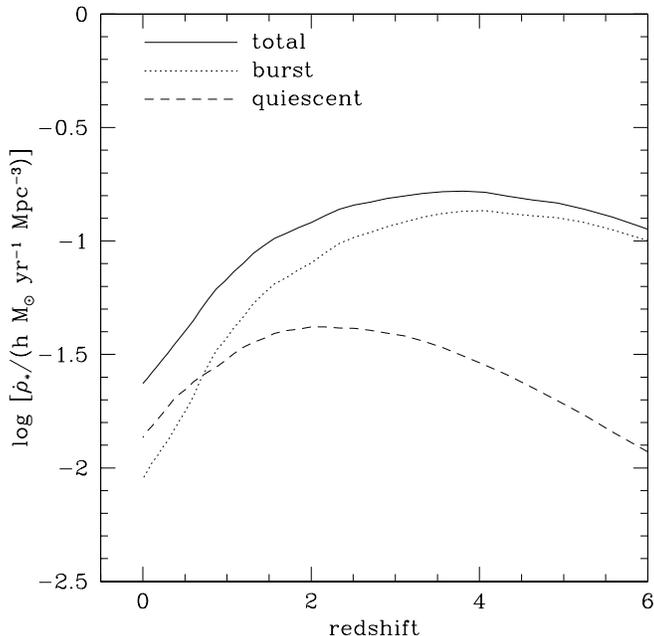,height=9truecm,width=9truecm}}
\caption{The star formation rate in the collisional starburst model 
is shown broken down into the contribution from quiescent star
formation and bursts. The bursting mode dominates at high redshifts,
and the quiescent mode dominates at low redshifts. The cross-over
occurs at a redshift of about 0.8.  }
\label{fig:fburst}
\end{figure}
\begin{figure}
\centerline{\psfig{file=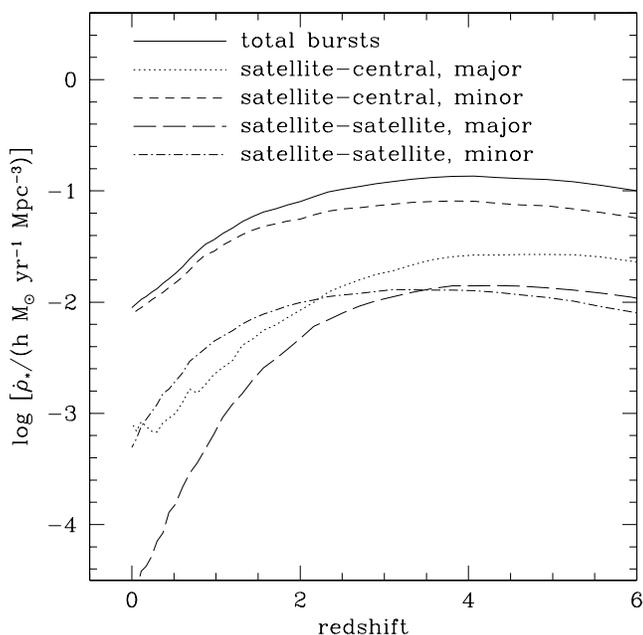,height=9truecm,width=9truecm}}
\caption{
Star formation in the bursting mode is shown broken down into
contributions from different kinds of merger events. The dotted and
dashed lines show contributions from major and minor mergers
(respectively) between a satellite galaxy and a central galaxy. The
long-dashed and dot-dashed lines show the contribution from major and
minor mergers between two satellite galaxies.  The dominant
contribution at all redshifts comes from minor mergers onto central
galaxies.}
\label{fig:burst_type}
\end{figure}

The star formation history of the Universe, as represented by the
popular ``Madau diagram'' \cite{madau:96}, is an important signature
of any scenario of galaxy formation. However, an accurate
determination of this diagram is far from straightforward: the
observations must be converted from luminosity to star formation rate
and corrected for incompleteness and dust extinction (see
Appendix). As new observations have been added at various redshifts,
the diagram has undergone a continuing metamorphosis from its original
form. Fig.~\ref{fig:new_madau_nodust} shows the star formation history
obtained from our burst and quiescent models, along with a compilation
of recent observations, where the observations have been corrected for
incompleteness but \emph{not} for dust extinction. The observations
have been converted to our $\lcdm$ cosmology as described in the
Appendix. Note that the constant efficiency quiescent model shows the
characteristic peak at $z\sim 1.5$ and the relatively steep decline at
higher redshifts that were notable features of the original Madau
diagram \cite{madau:96}.  However, the data points in that diagram at
$z=2.75$ and $z=4$ were based on the very small HDF survey and relied
on the Lyman-break technique to select galaxies in the desired
redshift range. As Madau et al. correctly emphasized, these points
should be regarded as lower limits. More recent observations based on
much larger, ground-based surveys with spectroscopic redshifts
\cite{steidel:99} show a much shallower evolution between $z\sim3$ and
$z\sim4$.

Another important recent observational development is a more convincing
estimate of the effects of dust extinction on the values derived from
luminosities in the far-UV. This has also produced a significant
modification of the diagram, as recently noted by
\citeN{steidel:99}. Fig.~\ref{fig:newmadau_dust} shows a subset of the
observations with a correction for dust extinction based on the
observational results (see Appendix for details). The CE quiescent
model is now seen to be highly inconsistent with the revised
observational estimates at high redshift.  The collisional starburst
model shows a gentler rise and a broad peak around $z\sim4$, with 
near-constant star formation density out to a redshift of $\sim 6$. This
appears very consistent with the recent optical estimates
after correction for dust extinction (see Appendix);
it is also consistent with the luminosity densities implied by the
far-IR and sub-mm sources detected at intermediate and high redshift
\cite{flores:99,hughes:98}. The redshift dependence of the star formation 
rate density in the accelerated quiescent model is very similar to
the collisional starburst model, although the overall star formation
rate density is a bit lower than the data.

Fig.~\ref{fig:fburst} shows the contribution to the star
formation rate density from the burst and quiescent modes separately
for our collisional starburst model. The bursting mode dominates at
high redshift but declines more steeply than the quiescent mode as
redshift decreases. A crossover therefore occurs, at $z\sim0.8$, and
quiescent star formation dominates the low-redshift Universe.

Fig.~\ref{fig:burst_type} shows the contribution to star formation in
the burst mode from different kinds of merger events. We show the
contribution from major and minor mergers between satellite and
central galaxies, and major and minor mergers between two satellite
galaxies. Note that in previous semi-analytic models, starbursts were
included only in the first case, i.e., satellite-central/major
mergers. We can see from the figure that the contribution from other
kinds of mergers, particularly satellite-central/minor, is quite
important.  The relatively small contribution of satellite-satellite
mergers may be an artifact of our rather questionable simplifying
assumption that all new cold gas is accreted onto the central galaxy;
this assumption tends to deprive small satellites of the fuel needed
to produce bright starbursts, but may not be realistic.

\subsubsection{Fossil Evidence}
\label{sec:results:fossil}
\begin{figure}
\centerline{\psfig{file=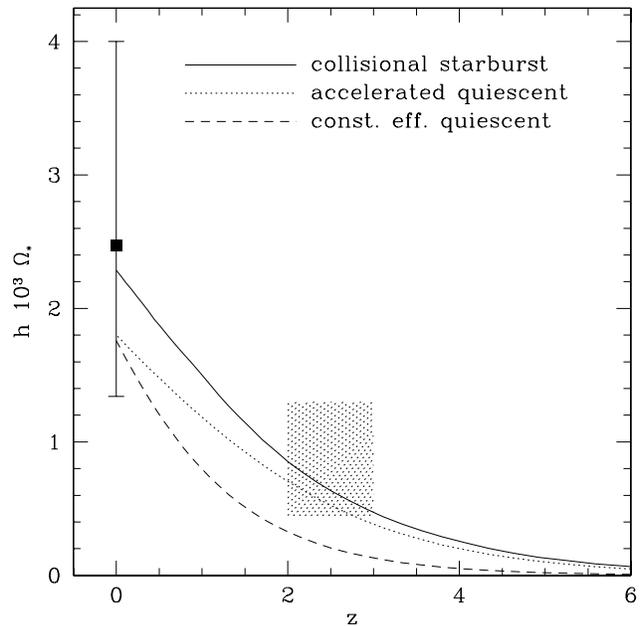,height=9truecm,width=9truecm}}
\caption{
The density in stars, in units of the critical density
($\Omega_{*}$). The observational estimate at $z=0$ is shown by the
symbol with the errorbar.  The observational constraint at high
redshift from ``fossil evidence'' (see text) is shown by the shaded
box. The collisional starburst model and the accelerated quiescent
model produce a significant fraction of their stars at high redshift,
wheras the constant efficiency quiescent model produces stars too
late. }
\label{fig:omega_star}
\end{figure}
Another way of constraining the star formation history of the Universe
is from the ``fossil evidence'' contained in the ages of stars in
present-day galaxies. \citeN{renzini} has argued that constraints on
the ages of early-type galaxies in clusters from the small observed
scatter of Fundamental Plane and colour-magnitude relations, combined
with the fraction of today's stellar mass contained in early-type
systems (bulges and ellipticals), can be used to deduce that one third
of the stars we see today must have formed at $z \ga
3$. \citeN{vandokkum:98} find that the formation redshift of
early-type galaxies in clusters may be relaxed to $z \ga 2$ in
cosmologies with a large cosmological constant (such as our \lcdm\
cosmology) because of the longer time that has ellapsed in such models.
Fig.~\ref{fig:omega_star} shows the density of stars in units of the
critical density for the burst and quiescent models. The point with
the large error bar at redshift zero is from the ``baryon budget'' of
\citeN{fhp}. All three models agree with this estimate within the
errors. The shaded patch shows the constraints on $\Omega_{*}$ at high
redshift from a Renzini-like argument, where we have simply divided
the stellar baryon budget at $z=0$ (and its error bar) by a factor of
three. The CE quiescent model does not produce enough stars at high
redshift, but the collisional starburst and accelerated quiescent
models meet this constraint. The CE quiescent model produces one-third
of the stars by $z=1.5$, and half of the stars by $z=0.9$. The
collisional starburst and accelerated quiescent models produce
one-third of their stars by $z=2.2$, and half of their stars by
$z=1.5$.

\subsubsection{Cold Gas Density}
\label{sec:results:omegagas}
\begin{figure}
\centerline{\psfig{file=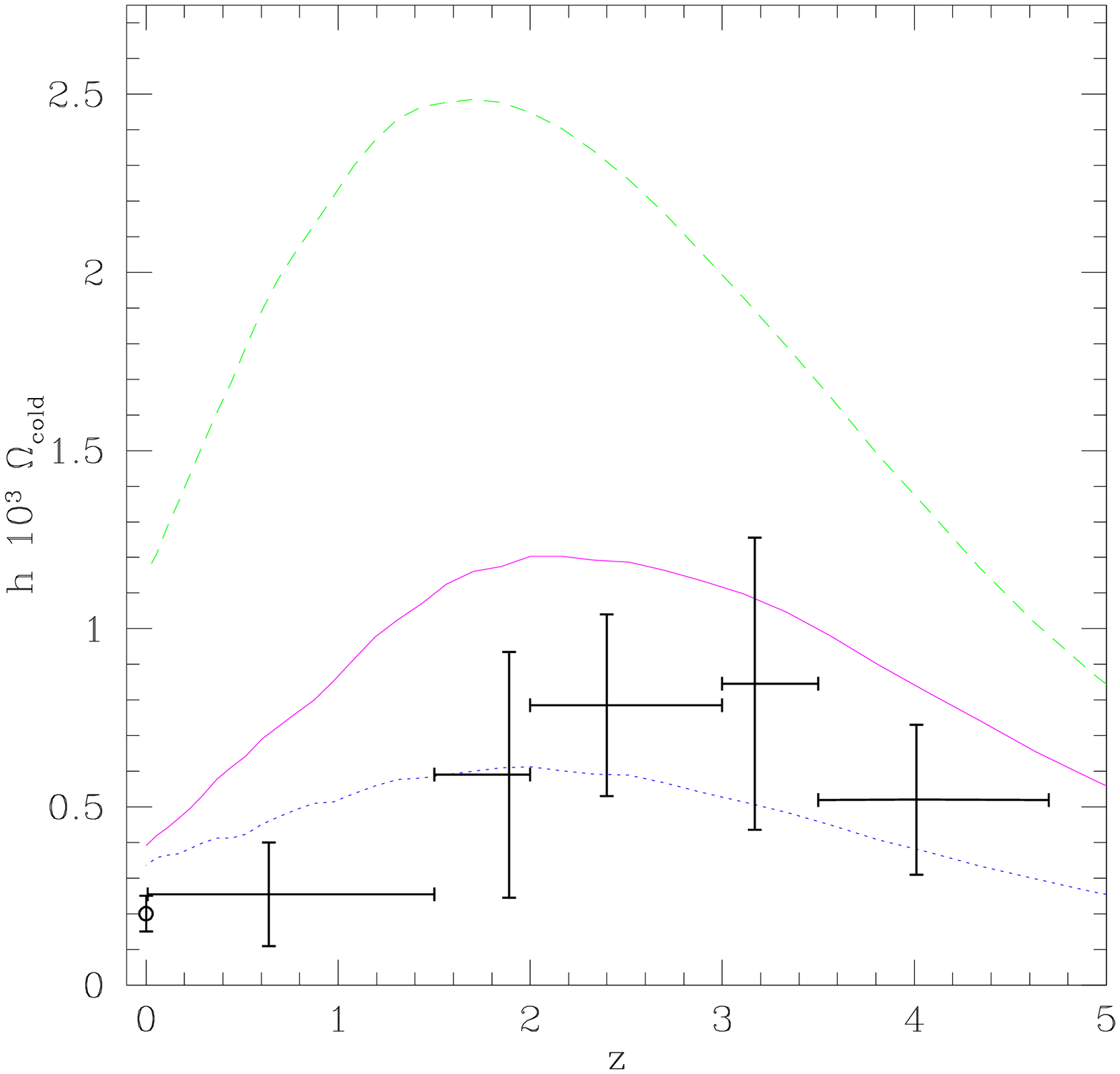,height=9truecm,width=9truecm}}
\caption{The evolution of \protect\omegagas\, as a function of
redshift.  Data points show the matter density in \protect\HI\, from
observations of DLAS
\protect\cite{storrie-lombardi:96}. The data point at $z=0$ is from local
\protect\HI\, observations \protect\cite{zwaan}. The quantity plotted for 
the models (line types are as in Fig.~\protect\ref{fig:omega_star}) is
generic ``cold gas'', some of which may be in the form of molecular or
ionized hydrogen, or may be in dust extinguished or low-column density
systems that would not be detected as damped systems. Thus the model
lines are upper limits on the quantity that is actually determined
from the observations ($\Omega_{\rm HI}$ in DLAS) and should lie above
the data points. Line types are the same as in 
figure~\protect\ref{fig:new_madau_nodust}.}
\label{fig:ogasz}
\end{figure}
Observations of quasar damped Lyman-$\alpha$ systems
(DLAS) provide an estimate of the
\HI\, and metal content of the Universe from $z\sim0.7$ to
$z\sim4$. Fig.~\ref{fig:ogasz} compares the models to an estimate of
the fraction of the critical density in the form of cold gas from the
observations of \citeN{storrie-lombardi:96} (the data converted to the
\lcdm\ cosmology was kindly provided by R. McMahon and C. Peroux).
The observations shown should be considered a lower limit on the total
mass of cold gas for several reasons.  Some dusty DLAS may also be
missed because their background quasars would be too dimmed to be
included in an optically selected, magnitude limited sample
\cite{pei-fall:95}. In addition a significant amount of mass could be
in the form of lower column density \HI\ clouds or in the form of
molecular or ionized hydrogen.

We can see from Fig.~\ref{fig:ogasz} that the CE quiescent model
(dashed line) overproduces the amount of cold gas by a factor that is
perhaps a bit large to be explained by these effects, whereas the
collisional starburst model (solid line) is more consistent with the
data.  The accelerated quiescent model (dotted line) is roughly
1$\sigma$ lower than the data at $z\ga 2$, and this may be a concern
since all of the effects mentioned above will tend to increase this
discrepancy. Note that the constraint provided by the DLAS is
complementary to that provided by observations of LBGs. We could
decrease the consumption of cold gas by decreasing the efficiency of
star formation $\tau^0_{*}$, but this would in turn produce fewer
bright LBGs.

\subsubsection{Metals}
\label{sec:global:metals}
\begin{figure}
\centerline{\psfig{file=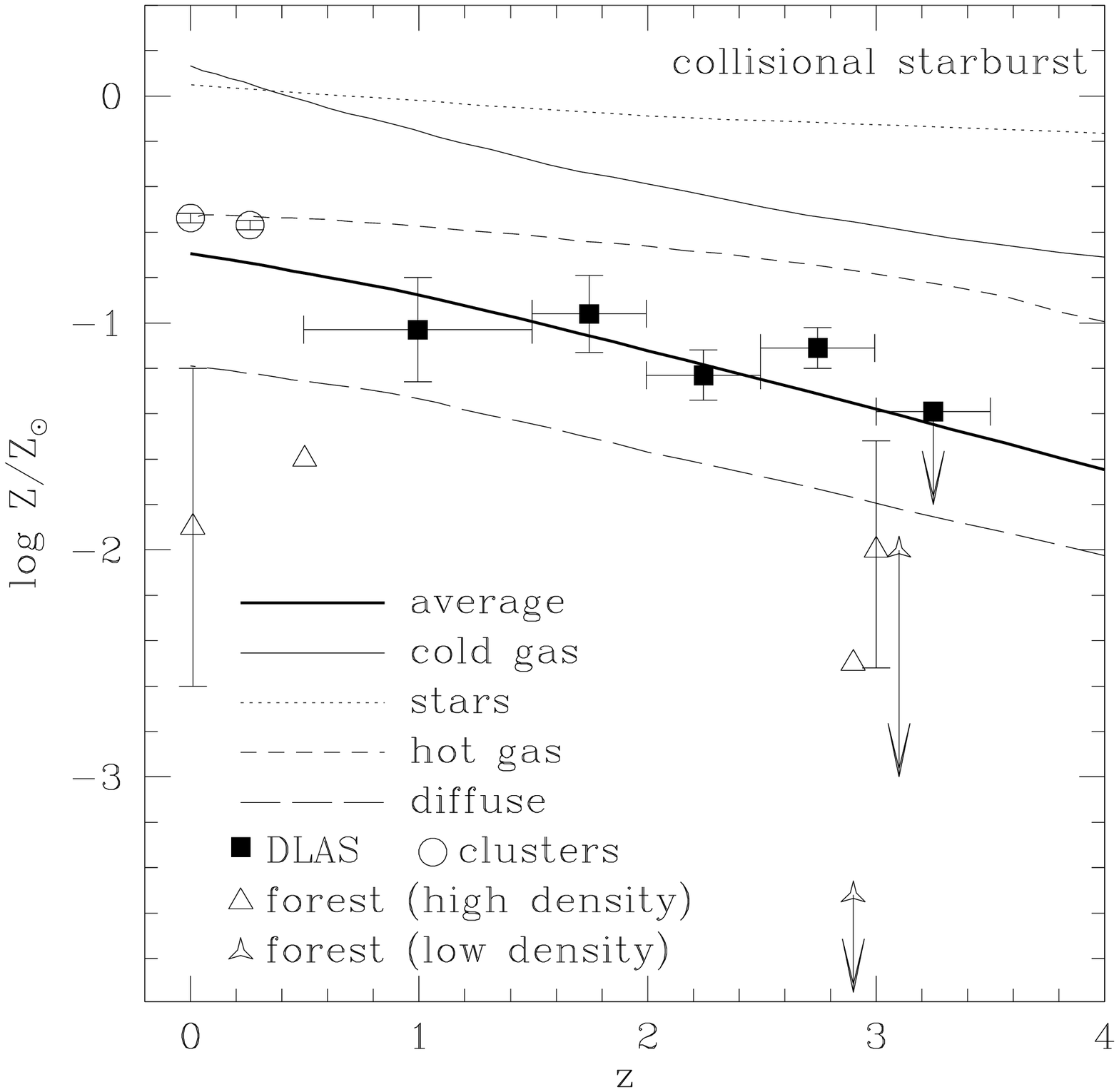,height=8truecm,width=8truecm}}
\caption{Metallicity as a function of redshift. The square symbols are 
from the measurements of Zn abundance in DLAS ([Zn/H$_{\rm DLA}$])
from \protect\citeN{pettini:dlas}. The filled dot at $z\sim0.3$ is
from measurements of Fe abundance in hot X-ray gas in clusters
\protect\cite{mushotszky:97}, and the dot at $z=0$ is the Fe abundance
from local clusters \protect\cite{yamashita:92,butcher:95}. Triangles
show estimates of metallicity in the diffuse IGM from observations of
the Lyman-$\alpha$ forest ($z\sim0$: \protect\citeNP{shull:98}; 
$z\sim0.5$: \protect\citeNP{bt:98}; $z\sim3$: \protect\citeNP{rauch:97}, 
\protect\citeNP{sc:96}, \protect\citeNP{lu:98}, \protect\citeNP{tytler:94}).
Lines show the collisional starburst model predictions of the
mass-weighted mean metallicity in the form of stars, cold gas, hot
gas, or diffuse (photoionized) gas, as labelled on the figure. }
\label{fig:metz_burst}
\end{figure}
\begin{figure}
\centerline{\psfig{file=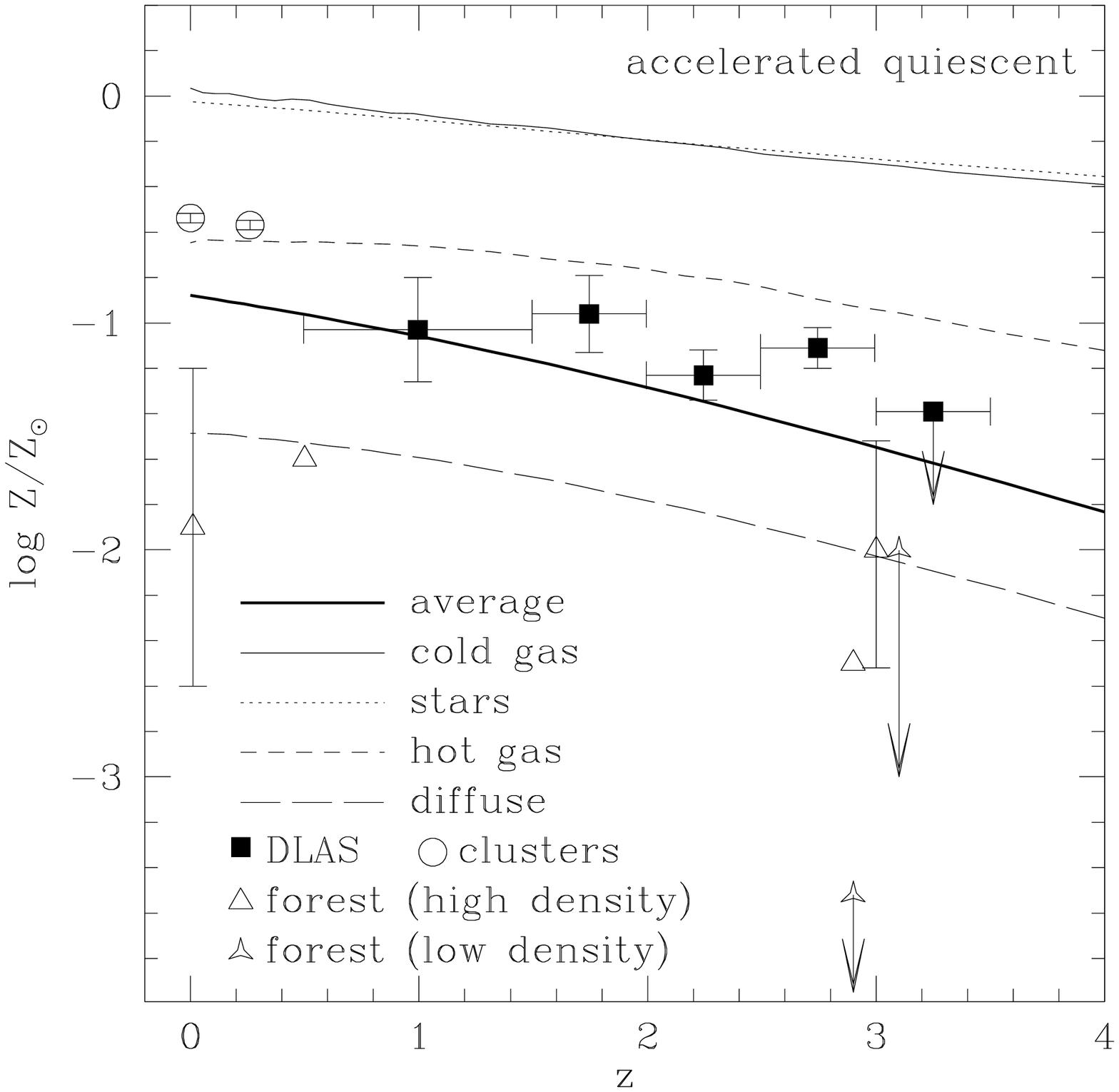,height=8truecm,width=8truecm}}
\centerline{\psfig{file=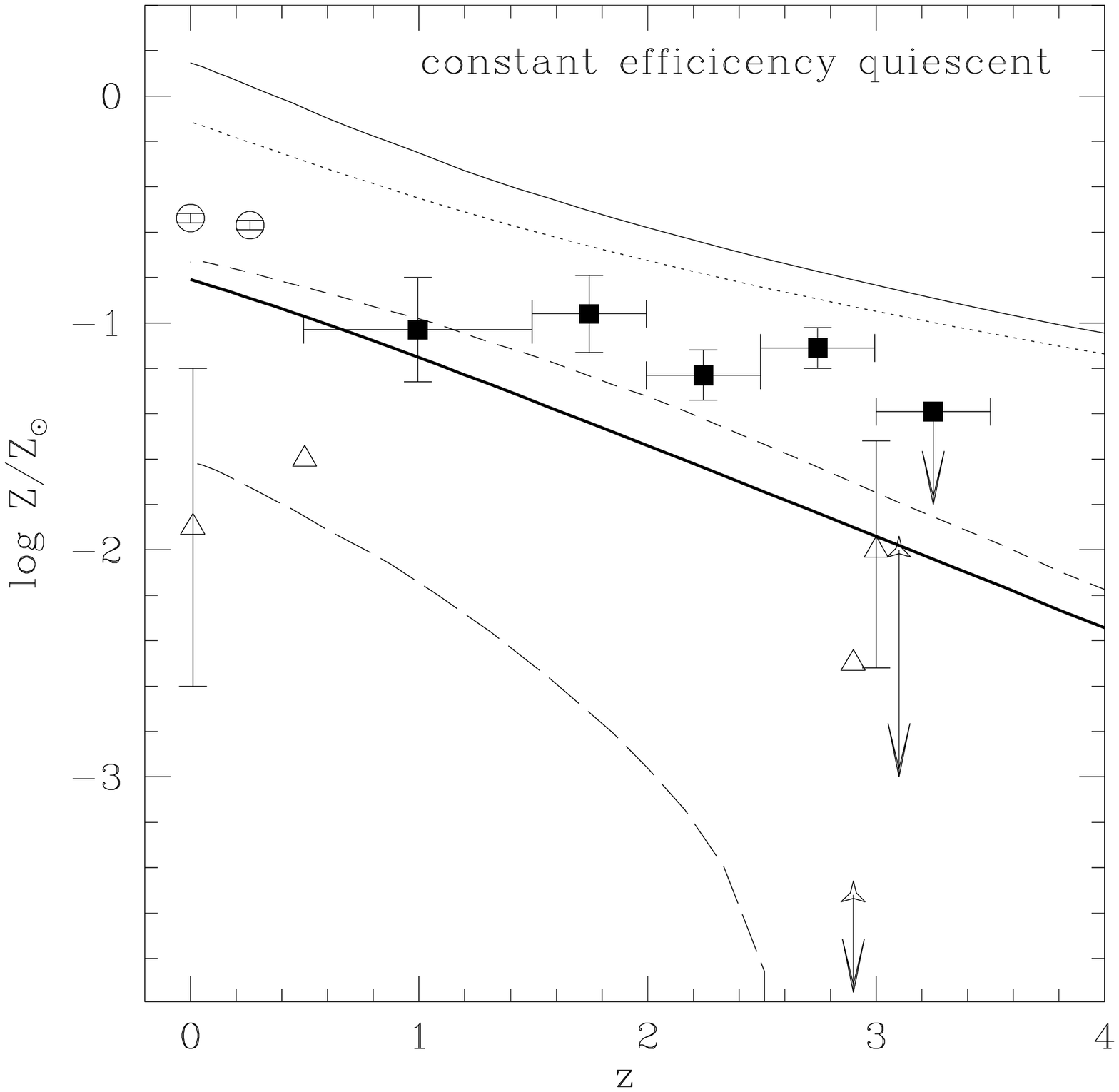,height=8truecm,width=8truecm}}
\caption{Same as Fig.~\ref{fig:metz_burst}, for the quiescent model. }
\label{fig:metz_quiescent}
\end{figure}
The metal content of the Universe is yet another tracer of the star
formation history.  The mean metallicity of the DLAS can be estimated
from Zn$_{\rm II}$ and Cr$_{\rm II}$ absorption lines, which are not
much depleted by dust \cite{pettini:dlas}.  The metallicity of
hot X-ray gas in clusters has been measured at $z\sim0.3$
\cite{mushotszky:97} and $z=0$ \cite{yamashita:92,butcher:95}. 
The metal content of the diffuse IGM is estimated from observations of
the Lyman-$\alpha$ forest (references in figure
caption). Figs.~\ref{fig:metz_burst} and~\ref{fig:metz_quiescent} show
the model predictions for the average metallicity of the entire
Universe (total mass in metals divided by total mass of gas), and for
the metallicity of stars, cold gas, hot gas in haloes, and diffuse
photoionized gas that has been ejected from haloes. The cold gas lines
from the models may be compared to the observations of DLAS, the hot
gas lines to the observations of hot gas in clusters, and the diffuse
gas lines to the metallicity of the higher-column-density
Lyman-$\alpha$ forest.  Note that the metals ejected from haloes in
our models most likely will not escape to large distances, but will
remain in the vicinity of the haloes. Therefore our predictions for
the metallicity of the ``diffuse'' gas are probably more appropriately
compared with the observational measurements in the higher column
density Lyman-$\alpha$ forest.

The differences among the model results are not large.  Overall, the
collisional starburst model predicts the highest metallicities, the
constant efficiency quiescent model the lowest.  Aside from the DLAS
metallicities, which are higher than the observations in all three
models (see below), the largest discrepancy is in the
Lyman-$\alpha$-forest, whose metallicity is too low in the constant
efficiency quiescent model.  The collisional starburst model predicts
the flattest metallicity evolution in all components, implying that
the average metallicity of stars at $z\sim3-4$ is already close to
solar.  The very small predicted evolution from $z=0$ to $z\sim 0.3$
in the metallicity of the hot gas (residing mainly in large,
cluster-sized haloes) in all three models is in good agreement with
the observations of hot gas in clusters
\cite{mushotszky:97}.

As noted, the metallicity of cold gas in all models is
systematically higher than that derived from DLAS. Since
the metallicities of the cold gas and stars in galaxies are tightly
coupled, it is difficult to imagine how one could make the metallicity
of the cold gas much lower yet still match the observed solar
metallicities of bright galaxies today. 
On the other hand, metallicities estimated from
DLA surveys may systematically underestimate the true values for
several reasons. First, dusty high-metallicity systems might dim any
quasar in the line of sight enough to discourage observers from trying
to obtain its spectrum \cite{pei-fall:95}. Second, the metallicities
of the outermost regions of local galaxies are often significantly
smaller than their central metallicities. By a simple cross-section
argument, there will be many more lines of sight passing through these
metal-poor outskirts than the metal-rich inner regions of the
galaxy. Finally, in the collisional starburst model, galaxies that
have experienced a recent burst of star formation, accompanied by
metal production, will have consumed nearly all of their cold gas and
will not produce damped systems in absorption.

An interesting side effect of the collisional starburst scenario is
that it predicts very early, efficient enrichment of the IGM and ICM
by supernovae driven winds. In our simple models, we have no
difficulty in ejecting enough metals to pollute the IGM to the
observed levels by $z\sim3$. This is consistent with the strong
metal-enriched outflows deduced from the spectra of observed LBGs
\cite{pettini:99}. This implies that a large fraction of the 
``missing metals'' 
at high redshift \cite{pettini:99b,pagel} may be in the form of hot
gas in clusters or in the IGM. Similarly, enrichment occurs early in
the accelerated quiescent model, again because galaxies with
relatively small masses have high star formation rates. By contrast,
in the constant efficiency quiescent model, most star formation takes
place in very massive haloes, and gas and metals cannot escape from
their deep potential wells. Significant pollution of the diffuse IGM
therefore occurs only at much later times ($z \la 1$).

\subsection{Properties of Galaxies at $z\sim3$}
\label{sec:results:properties}
Our results so far leave us with the conclusion that the constant
efficiency quiescent model is in serious conflict with the
observations. The best model seems to be the collisional starburst
model, although the accelerated quiescent model is not strongly ruled
out.  In this section, we concentrate on the collisional starburst
model, and investigate the predicted properties of $z\sim3$
galaxies in detail and compare them with the available
observations. For this purpose, we have created ``mock-HDF'' catalogs
with the same volume as the HDF. Dust extinction is included using our
usual recipe (see Section~\ref{sec:models:dust}).

\begin{figure*}
\centerline{\psfig{file=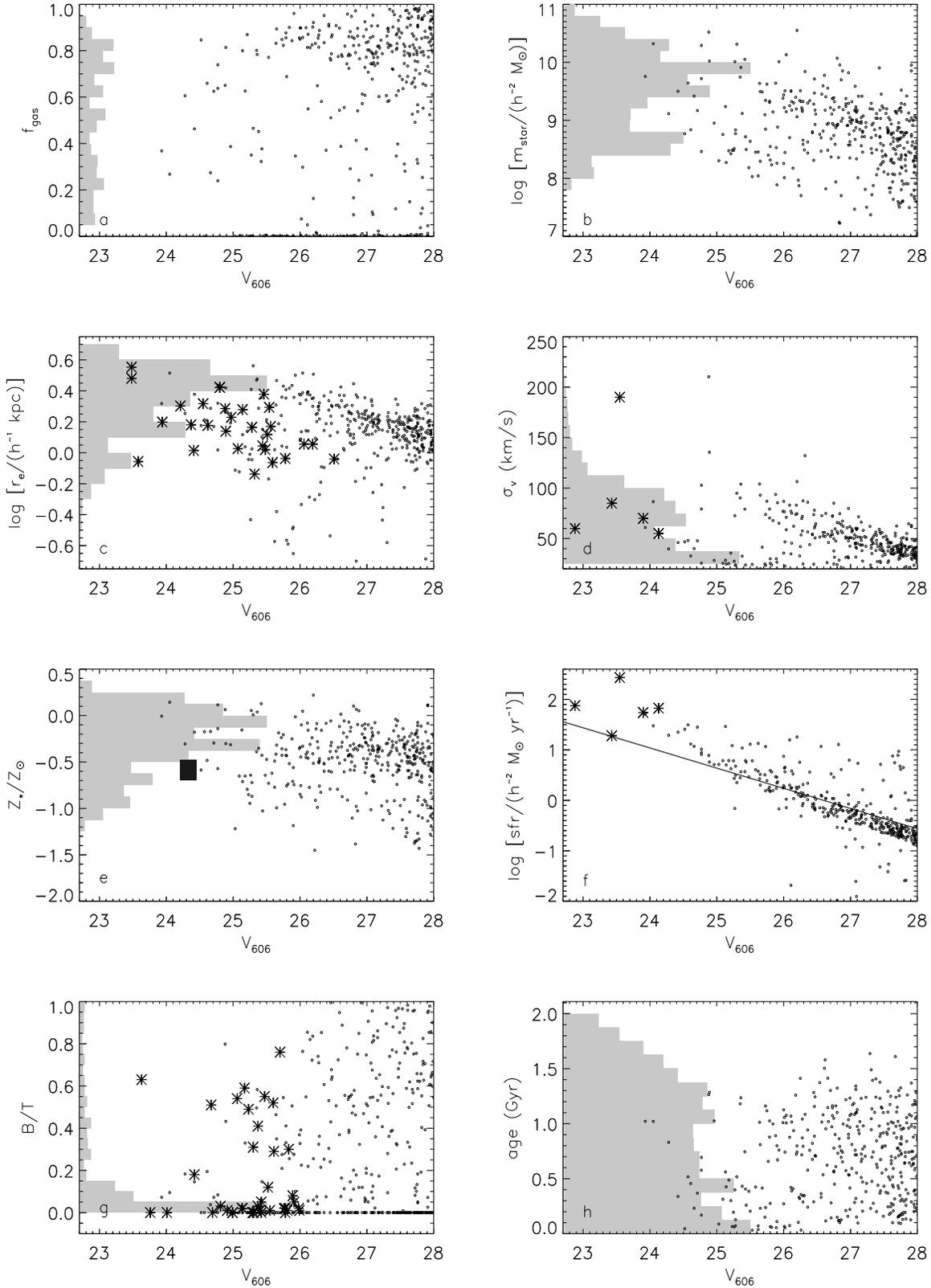,height=22truecm,width=16truecm}}
\caption{Small dots show various properties of galaxies in our
mock-HDF catalog, at $2 \leq z \leq 3.5$ as a function of the
$V_{606}$ (rest $\sim1500$ \AA) magnitude of the galaxy. The
histograms show the distribution of galaxies in this redshift range
that are brighter than $V_{606}=25.5$. The quantities shown are: a)
gas fraction, b) stellar mass, c) half-light radius, d) internal velocity
dispersion, e) stellar metallicity, f) star formation rate, g)
bulge-to-total mass ratio, h) mass-weighted mean stellar age. Large
star-shaped symbols and the dark square in panel e) show observational
estimates of some of these quantities (see text). }
\label{fig:galprops}
\end{figure*}

Fig.~\ref{fig:galprops} shows the properties of galaxies in the
redshift range $2 \la z \la 3.5$. The small dots show results as a
function of the $V_{606}$ (rest $\sim1600$ \AA) magnitude, for
galaxies in a typical mock-HDF catalog. Model galaxies have been
selected with a flat selection function over the redshift range $2 \la
z \la 3.5$. The histograms show the distribution of the same
quantities for galaxies brighter than $V_{606}=25.5$, to compare with
ground-based samples.  Because of the small volume of the HDF, we
average over many random realizations to obtain the histograms.

Panel a) shows the cold gas fraction $f_{\rm gas} \equiv m_{\rm
cold}/(m_{\rm cold}+m_{\rm star})$. The distribution of $f_{\rm gas}$
is nearly uniform over the full range, with a median value of $\sim
0.5$, and with many galaxies having values as high as $f_{\rm
gas}=0.8-0.9$. This is significantly higher than typical values for
local disc galaxies, which are around $0.1-0.25$ \cite{deblok:96}. We
do not have direct observational measures of the gas content of the
LBG population, but the high gas fractions we obtain are consistent
with the factor of $\ga 3$ increase in the total cold gas content of
the Universe ($\Omega_{\rm cold}$) from $z=0$ to $z\sim3$ from
observations of damped Lyman-$\alpha$ systems, as demonstrated
in Section~\ref{sec:results:omegagas}.

Panel b) shows the stellar masses of model LBG galaxies. Note that
stellar masses range from $10^8$ to $10^{11}$ h$^{-2} \msun$, on
average one or two orders of magnitude smaller than the stellar masses
of present-day $L_{*}$ spirals and ellipticals. Thus,
according to the collisional starburst model,
observed LBGs are not the
\emph{already fully assembled} progenitors of present day $L>L_{*}$
galaxies \cite{steidel:96a,giavalisco:96}.  It may well be the case
that most of the stars in these LBGs would end up within bright galaxies
at $z=0$; however, there will be several generations of merging in the
interim, and no simple one-to-one correspondence between the two
populations.  Accurate stellar masses for real LBGs are not yet
available; however, the model stellar masses that we obtain are very
compatible with the values calculated by
\citeN{sawicki:98} on the basis of multi-band photometry of galaxies 
in the HDF.

Panel c) compares the baryonic half-mass radii of model galaxies to
actual data (see Section~2.3 of SP for a discussion of how sizes are
estimated in our models; see \citeN{mmw:99} for more detailed
modelling of galaxy sizes). Typical values for the mock-catalog
galaxies are about a factor of two smaller than those of nearby bright
galaxies, further evidence that LBGs in the collisional starburst
model are not fully assembled $L_{*}(z=0)$ galaxies. The radii of the
model galaxies are in good agreement with the average half-light radii
of LBGs observed in the HDF (starred points;
\citeNP{giavalisco:96}, \citeNP{lowenthal:97}). Model radii at
$z=0$ are also in good agreement with the sizes of local galaxies, as
we showed in SP.  The significance of this agreement is less clear
here, however, since the half-light radii, particularly in the UV, may
well be considerably smaller than the baryonic half-mass radii that we
model.  However, radii measured in the observed near-IR (rest
$\sim$4000 \AA) from the NICMOS HDF are typically quite similar to
those shown here
\cite{dickinson:nicmos}.
 
Panel d) shows observed and model linewidths.  The velocity
dispersions of observed LBGs can be estimated based on the widths of
emission lines such as H$\beta$ or O$_{\rm III}$. Emission lines have
been detected for a few of the brightest LBGs, and the velocity
dispersions derived from the observed line-widths are rather small:
$50-90 \,\kms$ for four objects, and $180\,\kms$ for one object
\cite{pettini:98}. The widths of the observed emission lines probably
reflect the mass within about one effective radius. At such small
radii, there is probably little contribution from dark matter. We
therefore estimate the 1-D velocity dispersion in our models using the
expression
\begin{equation}
\sigma^2 = \frac{Gm}{c r_e} \, ,
\end{equation}
where $m$ is the total baryonic mass (cold gas and stars), $r_{e}$ is
the effective radius (we use the baryonic half-mass radius), and $c$
is a geometry-dependent factor \cite{phillips:97}, which we take to
equal 6, corresponding to a hot component with a density $\rho \propto
r^{-3}$ \cite{binneytremaine}. Although there is considerable
uncertainty as to how to model the linewidths of LBGs, we obtain good
agreement with the limited data available (\citeNP{pettini:98}; shown
by the large star symbols).

In panel a), we remarked on the rather small masses of galaxies in the
collisional starburst model.  That is paralleled here by the modest
velocity dispersions.  Masses (and velocities) tend to be small in
this model because small objects are elevated to the level of
visibility by starbursts.  Such an effect occurs to a lesser extent in
the accelerated quiescent model, but not at all in the constant
efficiency quiescent model.  This suggests that size indicators of all
types (masses, radii, velocity dispersions) are a way to discriminate
among models.  In the accelerated quiescent model, for example, the
shape of the distribution of dispersions is different --- it is
flatter and skewed towards larger $\sigma_v$) but has a similar mean
--- while the CE quiescent model typically has the largest
dispersions, ranging from 100 to 220 km/s, with a mean of about 140
km/s.  More detailed modelling and additional data will provide
important information on the baryonic masses of these objects.

Panel e) presents metallicities for model galaxies.
Little is known observationally about observed LBG metallicities,
although wide variations in the strength of C$_{\rm
IV}$ absorption (see \citeNP{steidel:96b}, \citeNP{lowenthal:97},
\citeNP{trager:97}) suggest that stellar metallicities have a broad
range of values. The stellar metallicities of our model galaxies are
consistent with this, ranging from about one-tenth solar to solar,
with a mean of about one-half solar.  This is also quite compatible with
the estimate based on interstellar absorption lines in the lensed
galaxy MS 1512-cB58 (\citeNP{pettini:99}; shown as the dark square on
panel e). A weak metallicity-luminosity relationship is already in
place at $z=3$, with the brightest objects typically showing higher
metallicities.

A tight correlation between rest-UV luminosity and 
instantaneous star formation rate is often assumed in order to
estimate the former from the latter. The solid line 
in panel f) shows the relation
of this sort used by \citeN{mpd:97}. However, galaxies with complex
star formation histories, particularly episodic ones, may show a
non-negligible scatter from this relationship, as shown by our model galaxies.
Dust extinction also increases the scatter in this relation
and moves the brightest, most extinguished galaxies off of it. The
large star symbols in this panel show galaxies from the sample of 
\citeN{pettini:98}, where the star formation rates have been estimated from
H$\beta$ emission lines.  Star formation rates for the brightest
galaxies in the collisional starburst model approach 100 solar masses
per year, in agreement with observed values. The quiescent models
predict far fewer galaxies with such high star formation rates, as
reflected in Fig.~\ref{fig:lf_z3_burst} and \ref{fig:lf_z3_quiescent}.

Panel g) shows the mass ratio of stars in the bulge to the total
stellar mass ($B/T$) for the model galaxies. We find that bright
galaxies are biased towards low $B/T$ ratios, indicating that the
galaxies are disc dominated. Observed $B/T$ values (starred points)
are also shown for bulge-disc
decompositions of HDF galaxies, using the procedure described in
\citeN{marleau:98}, where ``bulge'' and ``disc''
components were determined by fitting observed surface brightness
profiles with a S\'{e}rsic or an exponential form. The availability of
photometric redshifts allowed us to select galaxies in the redshift
range of interest ($2.5 \le z \le 3.5$). These observations were
compiled and provided to us by L. Simard and K. Wu. The resulting
distribution of $B/T$ with $V_{606}$ magnitude looks very similar to
the predictions of our models, including the weak apparent correlation
of magnitude with $B/T$.  However, our $B/T$ values are weighted by
mass, while observed values are weighted by light.  This difference
might bias the measured values in some unknown way.  It is also
interesting to note that about 20 percent of the HDF galaxies
(brighter than $m_{AB}(1500) \sim 26$) have $B/T > 0.40$, similar to
the fraction of early-type galaxies found in the nearby Universe.

The presence of an exponential profile does not mean that a
classical disc is present, and similarly an observed $r^{1/4}$ profile
does not imply the existence of an early-type galaxy in the classical
sense. It is clear that traditional morphological definitions must be
expanded and modified to usefully discuss the inhabitants of the
high-redshift Universe. Many LBGs show a centrally concentrated
$r^{1/4}$ core within a more diffuse envelope
\cite{giavalisco:96}, yet many also show pronounced substructure 
and signs of disturbance, even in the rest-visual NICMOS images
\cite{lowenthal:97,conselice:98}. 
The collisional starburst scenario predicts that a measureable
fraction of LBGs should show significant substructure and
morphological peculiarity and are quite unlikely to resemble classical
disc galaxies.  A quantitative statistical analysis of the morphology
of observed LBGs could place important constraints on the collisional
starburst scenario.  We intend to pursue this topic in future work
(preliminary results are presented in \citeNP{somerville:iap}).

Accurate stellar ages of LBGs would provide a good test of models.
However, the determination of ages from observed colours is
complicated by the degenerate effect of reddening due to dust.  Some
observational papers discussing the LBG population have suggested that
the total duration of star formation may be as large as 1~Gyr
\cite{steidel:96b,pettini:97}, based on the ${\mathcal R}-K$ colours of the
objects.  However, \citeN{sawicki:98} conclude that the dominant
stellar population of the LBGs is less than 0.2 Gyr old, with median
ages of $\sim10$-36 Myr.  This conclusion was based on their
comparison of model SEDs to photometric data from 5 filter bands
(VIJHK) spanning the Balmer break. The IR photometric data is helpful
in breaking the age-extinction degeneracy, although some degeneracy
remains. Panel h) in Fig.~\ref{fig:galprops}
shows mass-weighted mean stellar ages
for model galaxies at $z=3$ as a function of apparent magnitude
(note that the \emph{luminosity} weighted ages, which corresponds more
closely to the quantities estimated by \citeN{sawicki:98}, are
considerably younger; see \S\ref{sec:results:colors}). 
The distribution of ages is fairly flat, with many
objects having very young ages (indicating that they formed most of
their stars in a recent burst) and some having ages close to the age
of the Universe at the mean redshift of the sample \footnote{The age
of the Universe in this cosmology is 3.2 Gyr at $z=2$, 2.1 Gyr at
$z=3$, and 1.8 Gyr at $z=3.5$.}.

\begin{figure}
\centerline{\psfig{file=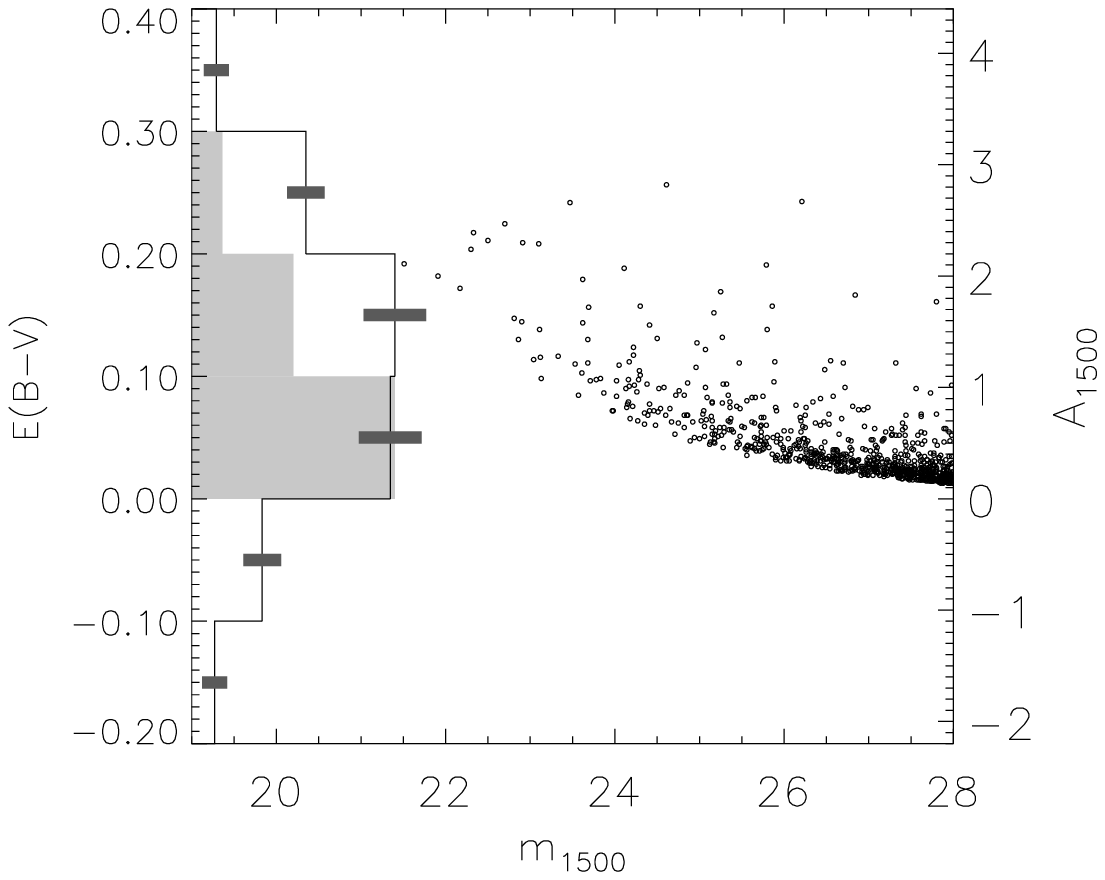,height=8truecm,width=9truecm}}
\caption{Extinction at rest 1500 \AA\ for
the model galaxies at $z=3$, as a function of their
\emph{unattenuated} UV magnitude (see text for details of the dust
recipe). The lefthand axis shows the extinction in terms of
(rest-frame) E(B-V), and the righthand axis shows it in terms of
magnitudes of extinction at 1500 \AA.  The scatter at fixed magnitude
is due to the random inclination chosen for each galaxy. The shaded
histogram shows the distribution of rest E(B-V) for the bright
(\emph{attenuated} $m_{1500} < 25.5$) galaxies in the models. The
unshaded histogram shows the distribution of rest E(B-V) for LBGs in
the ground-based sample to similar magnitudes by
\protect\cite{adelberger:00}.}
\label{fig:ebminusv}
\end{figure}

\begin{figure*}
\centerline{\psfig{file=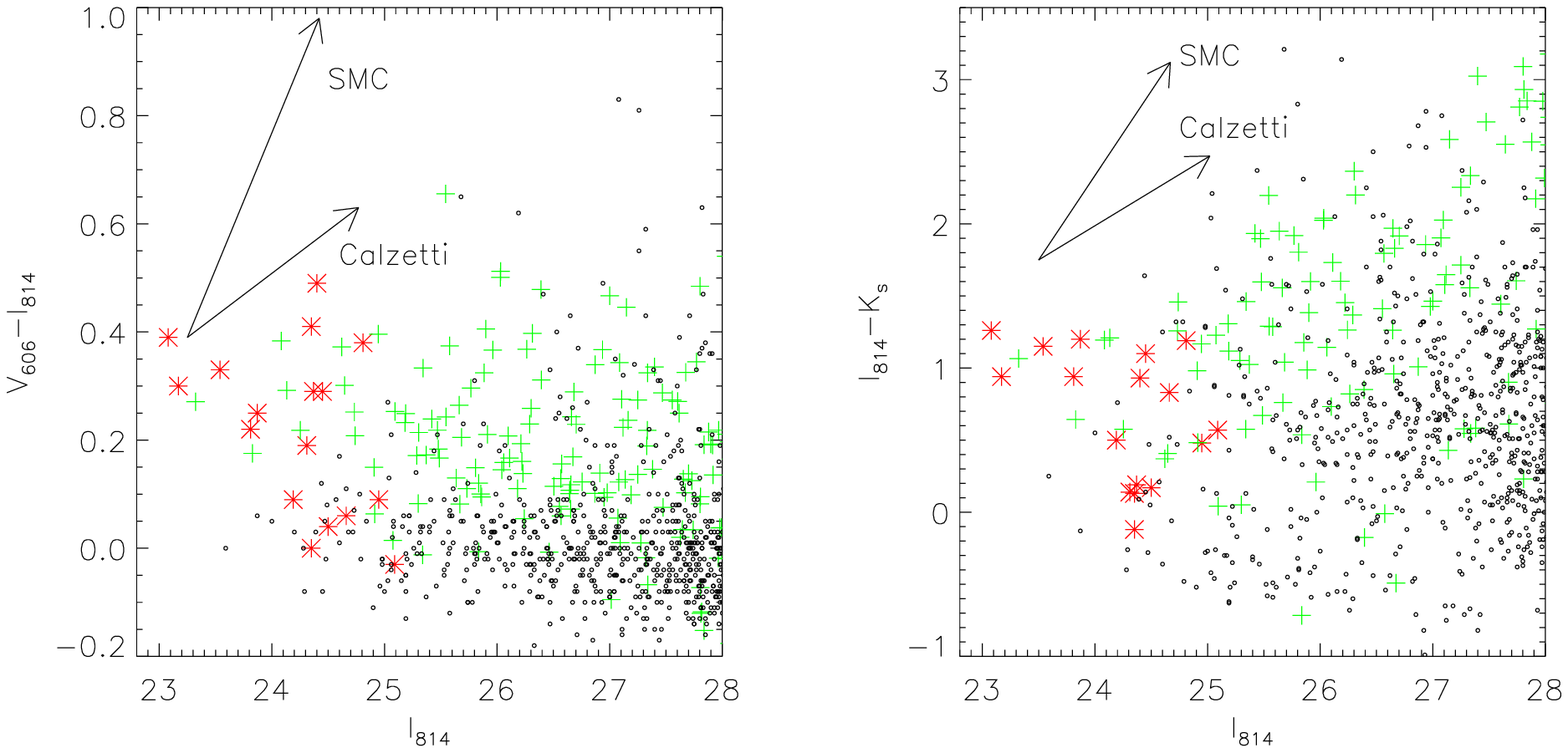,height=8truecm,width=16truecm}}
\caption{
Observed V$-$I (left) and I$-$K (right) colours for galaxies at
$z\sim3$. The plus symbols show the colours of galaxies in the
HDF with photometric redshifts in the range $2.0 \le z \le 3.5$
\protect\cite{chen:99a,lanzetta:99a}. Large star
symbols show colours of HDF galaxies with spectroscopic redshifts
\protect\cite{lowenthal:97,sawicki:98}. 
The small dots show the colours of model galaxies extracted from the
mock-HDF catalog, assuming a flat selection function in the range $2.0
\le z \le 3.5$. Dust extinction is included in the model galaxy
colours using our usual recipe. The arrows show the reddening vector
for an $L_{*}$ galaxy at $z=3$ assuming either a Calzetti or SMC
extinction curve. }
\label{fig:colormag}
\end{figure*}

\begin{figure*}
\centerline{\psfig{file=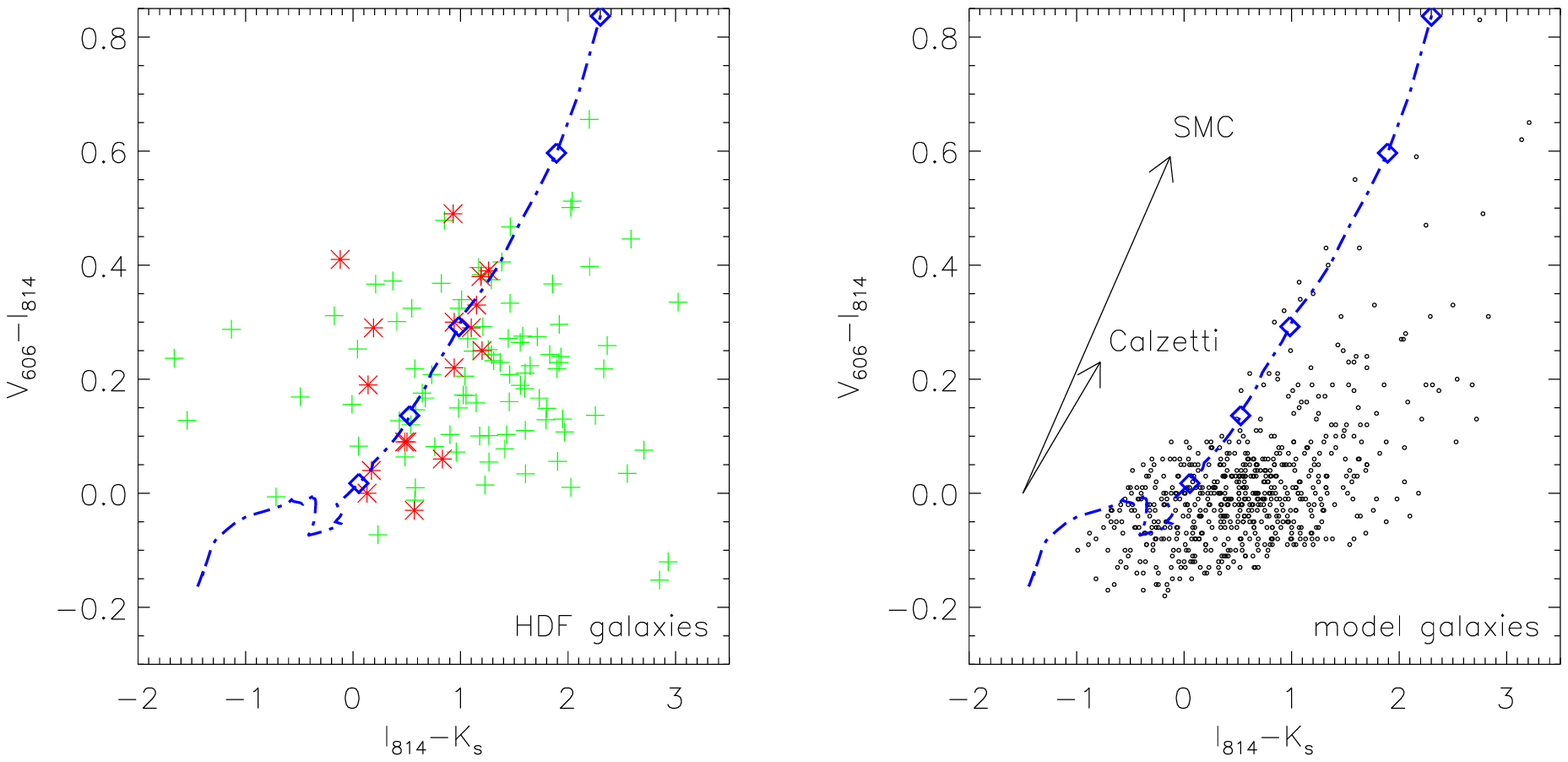,height=8truecm,width=16truecm}}
\caption{Two-colour diagram (observed V$-$I vs. I$-$K) for galaxies at
$z\sim3$. In the left panel, large star symbols and plus symbols show
the colours of HDF galaxies with spectroscopic and photometric
redshifts (respectively) in the range $2.0 \le z \le 3.5$ (references
in text). Small dots show the colours of model galaxies in the same
redshift range from the mock-HDF catalog (assuming a flat selection
function from $2.0 \le z
\le 3.5$), including dust extinction as usual. Dot-dashed lines show
the unreddened colours of instantaneous burst populations of various
ages from the solar metallicity Bruzual \& Charlot models (GISSEL) at
$z=3$. The five diamond symbols mark the colours of single-burst ages
of $\sim25$, 50, 100, 250, and 320 Myr, from lower left to upper
right. The arrows show the reddening vector for an $L_{*}$ galaxy at
$z=3$, assuming a Calzetti or SMC type extinction curve. The length of
the vector represents an extinction of a factor of five at 1500 \AA. }
\label{fig:viik}
\end{figure*}

\subsubsection{Dust Extinction and Colours}
\label{sec:results:colors}
As  described in Section~\ref{sec:models:dust}, we have assumed a
very simple recipe for dust extinction, in which the optical depth of
a galaxy is a deterministic function of its intrinsic UV
luminosity. Recall that the resulting extinction is also a function of
the galaxy's inclination, which we chose
randomly. Fig.~\ref{fig:ebminusv} shows the effects of dust
reddening on the model galaxies in our mock-HDF catalog as a function
of the unextinguished magnitude at rest 1500 \AA. We convert the
extinction in magnitudes (shown on the righthand axis) to a rest-frame
colour excess E(B-V) using the recipe given in
\citeN{calzetti:97a}. The solid histogram shows the distribution for
the brightest (\emph{attenuated} magnitude $m_{1500} < 25.5$) model
galaxies. For comparison, we also show the distribution of rest E(B-V)
(provided to us by K. Adelberger) obtained from the spectral slopes of
LBGs from the ground-based sample of
\citeN{steidel:99}, using the technique described by
\citeN{meurer:99}. 
A detailed discussion of dust extinction in observed LBGs is presented
in
\citeN{adelberger:00}. Note that negative values of E(B-V) are
unphysical and result in part from the presence of Lyman-$\alpha$
emission in some of the spectra (not included in our modelling). In
about half of the cases, Lyman-$\alpha$ is seen in absorption and this
broadens the E(B-V) distribution redwards. Within these uncertainties,
the distribution we obtain in our models is fairly similar to the
observed one. This gives us some confidence that our simple dust model
is at least roughly compatible with the best current observational
estimates.

We investigate the colours of our model galaxies, including the
effects of dust, using the same recipe discussed
above. Fig.~\ref{fig:colormag} shows the $V_{606}-I_{814}$ and
$I_{814}-K_s$ colours of galaxies in our mock-HDF (small dots) along
with galaxies with spectroscopic \cite{lowenthal:97,sawicki:98} and
photometric redshifts
\cite{chen:99a,lanzetta:99a}. It is immediately apparent that the model
galaxies are too blue in $V_{606}-I_{814}$, by about 0.2 magnitudes,
while the $I_{814}-K_s$ colours are in reasonably good agreement with
the data. One might be tempted to conclude from this that the
starburst model produces galaxies with too young a stellar population;
however, this is unlikely to be the explanation, as $I_{814}-K_s$
spans the Balmer break at this redshift and is a much better indicator
of the age of the stellar population. The $V_{606}-I_{814}$ colour
reflects the spectral slope in the far-UV ($\sim$ 1500--2000\AA), and, as we
have discussed, this is strongly correlated with the far-IR excess,
and hence with the amount of dust extinction
\cite{meurer:99}. This seems to indicate that we have underestimated the
amount of dust extinction in our model galaxies, or that the true
extinction curve is steeper than the Calzetti curve that we have
assumed throughout --- perhaps closer to an SMC curve. By comparing
the reddening vectors shown in the figure, one can see that, had we
used an SMC curve, both sets of colours would have been in quite good
agreement with the data. 

A similar conclusion may be drawn from
Fig.~\ref{fig:viik}. The left panel shows a two-colour diagram
($V_{606}-I_{814}$ vs. $I_{814}-K_s$) for the observed HDF galaxies
with spectroscopic (stars) and photometric (crosses) redshifts, as
before. It is interesting to note the four or five photo-z galaxies
with anomalously blue $I_{814}-K_s$ colours for their
$V_{606}-I_{814}$ colours. A cross-check reveals that these galaxies
are all fainter than $I_{814}=25.5$ and therefore are not bright
enough to be contained in the spectroscopic sample of
\citeN{lowenthal:97}. By comparing with the dot-dashed lines, which
show the unreddened colours of a single-age stellar population from
the GISSEL98 models (diamonds mark ages of $\sim25$, 50, 100, 250, and
320 Myr from bottom left to top right), we see that their
$I_{814}-K_s$ colours are characteristic of an extremely young (less
than 25 Myr) stellar population but the $V_{606}-I_{814}$ colours are
considerably redder --- perhaps these are very heavily extinguished
starbursts? The righthand panel shows where our model galaxies from
the mock-HDF lie in this diagram. We do not see these anomalous
objects in the models, even if an SMC extinction law is assumed, which
could be an indication that our dust modelling is too simplistic
(which would hardly be surprising), or that these objects are unusual,
have been assigned incorrect redshifts, or have large photometric
errors.

\section{Comparison with Previous Work}
\label{sec:previous}
\begin{figure}
\centerline{\psfig{file=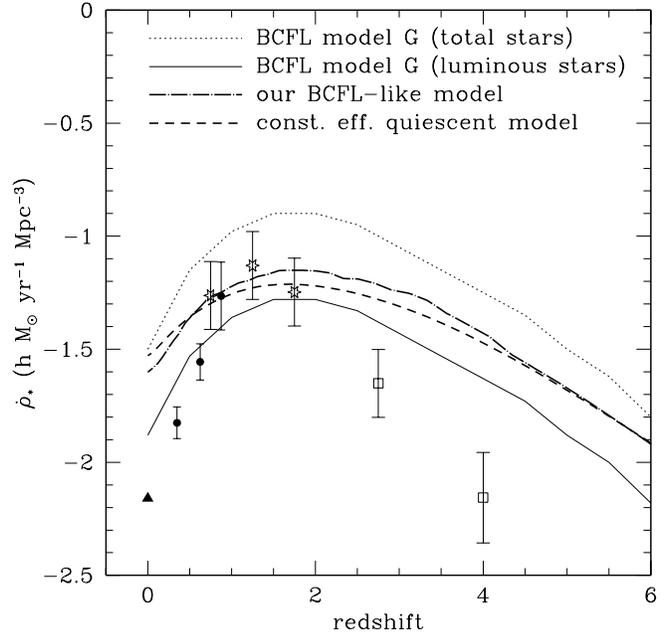,height=9truecm,width=9truecm}}
\caption{The (almost) original Madau diagram
\protect\cite{mpd:97}. The models shown are: model G of BCFL, our
attempt to reproduce model G of BCFL, and our constant efficiency
quiescent model. All three are seen to be similar. The dotted line is
the curve shown in Fig.~16 of BCFL. Our conversion of the data
points from luminosity to star formation rate is different than the
one used by BCFL, and is comparable to the star formation in
\emph{luminous} stars, shown for BCFL model G by the solid line. }
\label{fig:oldmadau}
\end{figure}

The results of this sort of modelling, particularly the redshift
evolution, can be quite sensitive to the details of the recipes used
to model the physical processes.  This explains some of the
differences between our results and those of other groups using
similar techniques.  For example, previous semi-analytic modelling of
high-redshift galaxies (BCFL) gave a rather different picture of the
star formation history of the Universe and the nature of the
Lyman-break galaxies. BCFL concluded that observable LBGs form only in
very massive haloes, $\ga 10^{12}
\hmsun$, and implied that the dominant mode of star formation at all
redshifts is quiescent. They state explicitly, and show in their
Fig.~7, that in their models ``most galaxies...never experience star
formation rates in excess of a few solar masses per year.'' This is
clearly quite different from the results of our best-fitting
models. These differences arise from the different ingredients that we
have chosen to include in our models. BCFL made several assumptions
that made the mass-to-light ratios of their haloes higher than
ours. Based on our attempts to reproduce their results, we believe
that some of the important differences are:
\begin{enumerate}
\item BCFL assumed primordial metallicity for all hot gas, reducing the
efficiency of gas cooling (see Fig.~2 of SP).
\item The explicit $V_c$ dependence in their star formation recipe
(Eqn.~\ref{eqn:sfr} of this paper; Eqn. 2.5 and 2.10 of CAFNZ) makes
star formation less efficient in small $V_c$ haloes. This suppresses
star formation at high redshift, when typical halo masses are smaller.
\item The strong $V_c$ dependence of their supernovae feedback recipe
(Eqn.~2.8 and 2.11 of CAFNZ) further suppresses star formation in
small haloes.
\item BCFL assume that only a fraction of the stars are luminous. 
In their fiducial model A, they assume that $f^{*}_{\rm lum}=0.36$
($\Upsilon=[f^{*}_{\rm lum}]^{-1}=2.8$, in their notation), and in
model G $f^{*}_{\rm lum}=0.42$. This results in stellar mass-to-light
ratios that are larger by a similar factor.
\item Although merger-driven starbursts are included in the models of BCFL, 
they occur only in major mergers between satellites and central
galaxies. This neglects the contribution from satellite-satellite
mergers and minor mergers, which we have shown can be quite
significant (see Fig.~\ref{fig:burst_type}).
\end{enumerate}

Fig.~\ref{fig:oldmadau} shows the star formation history of the
Universe as represented by the (almost) original Madau diagram
\protect\cite{mpd:97}, along with the
results of BCFL model G (the closest of their models to our fiducial
cosmology and IMF)\footnote{Note that in Fig.~ 16 of BCFL, the data
points shown were calculated using a different conversion from
luminosity to star formation than ours (see Table~\ref{tab:sfconv} in
our Appendix, and Table~5 of BCFL). In addition, BCFL plotted the
``total'' star formation rate in the models and corrected the
observations assuming that only a fraction $[\Upsilon]^{-1}$ of the
stars are luminous. We instead plot the star formation of
\emph{luminous} stars in all of the models.}. Also shown are our attempts to
reproduce their results by adopting the assumptions listed above. Many
differences in the details of the modelling remain, so it is not
surprising that our results do not agree exactly, but they are similar
enough to give us confidence that we understand the main effects. 

We also show the results of our constant efficiency quiescent model,
the same model shown throughout the previous sections. It turns out
that the results of model G of BCFL are very similar to those of our
constant efficiency quiescent model. Although the efficiency of
quiescent star formation is actually \emph{lower} at early times in
the models of BCFL (see points (ii) and (iii) above), starbursts in
satellite-central major mergers were included (see point (v) above)
and this contributes some additional bright galaxies.  One can see
from Fig.~\ref{fig:oldmadau} and can further verify by comparing the
luminosity functions shown in Fig.~\ref{fig:lf_z3_quiescent} and
Fig.~\ref{fig:lf_z4_quiescent} with Fig.~15 of BCFL, that, for our
purposes, the models produce very similar results. Like our CE
quiescent model, the models of BCFL produced just enough light and
just enough bright galaxies at $z\sim3$
\emph{if dust extinction was neglected}. 
We now see that, when dust extinction
is taken into account, these models produce a luminosity function that
falls off too steeply on the bright end, and predict a steep fall-off
in the number density of bright galaxies and the integrated luminosity
density with redshift. This results in very few bright galaxies at
redshifts of $\ga5$, in sharp contrast to the collisional starburst
model or the accelerated quiescent model.

As noted in Section~\ref{sec:models}, our accelerated quiescent star
formation recipe is effectively very similar to the one usually used
by the Munich group (e.g. KWG93). We have shown that this recipe
produces fairly good agreement with many of the observations, but does
not produce enough very bright LBGs and also underpredicts the amount of
cold gas at high redshift compared with observations of damped
Lyman-$\alpha$ systems.  In a recent paper, \citeN{kh} reached a
similar conclusion about the cold gas problem and adopted a similar
solution. They write their quiescent star formation recipe as
$\dot{m}_* = \alpha m_{\rm cold}/t_{\rm dyn}$. As noted before, the
dynamical time $t_{\rm dyn}$ scales approximately as $t_{\rm dyn}
\propto (1+z)^{-3/2}$, so their adopted scaling of the parameter
$\alpha$ with redshift as $\alpha(z) = \alpha(0) (1+z)^{-\gamma}$ with
$\gamma =$ 1--2 is very similar to our ``constant efficiency
quiescent'' recipe. They also include bursts in major
satellite-central mergers, and find that these bursts are responsible
for about two-thirds of the star formation at high redshift.  They
find that this model, which is similar to our favored collisional
starburst model, produces good results for the redshift evolution of
the space density of bright \emph{quasars}.  Taken together, this
suggests a picture in which the brightest inhabitants of the
high-redshift Universe, LBGs and quasars, may be produced by the same
process: strong inflows induced by mergers.

It is also interesting to compare our results to those of cosmological
$N$-body simulations including hydrodynamics and star
formation. \citeN{katz:99} performed a detailed study of the
clustering of high redshift galaxies in such simulations, and
concluded that the observed clustering and number densities could be
easily explained in a variety of CDM models. In this respect, their
results are quite similar to the massive halo models discussed in
Section~\ref{sec:intro}. However, although their simulations included
gas cooling and star formation, they assumed a monotonic relation
between baryonic mass and UV luminosity. As we have already discussed,
and as acknowledged by \citeN{katz:99}, the effects of episodic star
formation and dust could significantly complicate this relationship. A
more direct look at the nature of the $z\sim3$ galaxies in these
simulations was taken by \citeN{romeel:marseille}. Here, the star
formation history of the galaxies was extracted and convolved with
stellar population synthesis models to obtain estimates of their
luminosities, and the effect of dust was considered. They then find
that the luminosity function at $z\sim3$ is in agreement with the
observations when dust extinction is included. The galaxies brighter
than $25.5$ in their simulations have fairly large stellar masses (log
$m_{\rm star} \ga 10.4$ for their value of $h_{100}$=0.65) and contain
a significant older stellar population; i.e. they are not
predominantly young bursts. However, Dav\'{e} et al. acknowledge that
their simulations do not have sufficient resolution to properly model
the collisional starburst mechanism that we have invoked here, so they
cannot rule out the possibility that small starbursts could also be
present. It is also worth noting that in these simulations, the star
formation efficiency scales as the gas density to some power, which
will give a similar behaviour to our ``accelerated quiescent'' star
formation recipe, and this may partially explain why they obtain
sufficient numbers of LBGs without the need for an additional
population of starbursts. The volume of their simulations is extremely
small ($\sim 11 h^{-1}$ Mpc on a side), far too small to probe the
very bright end of the luminosity function, where we found the most
pronounced differences between the collisional starburst model and the
accelerated quiescent model.

\section{Dependence on Modelling Assumptions}
\label{sec:variations}
This section discusses the sensitivity of our results to the
cosmology, IMF, and stellar population modelling.

\subsection{Cosmology and Power Spectrum}
\label{sec:variations:cosmo}
\begin{figure}
\centerline{\psfig{file=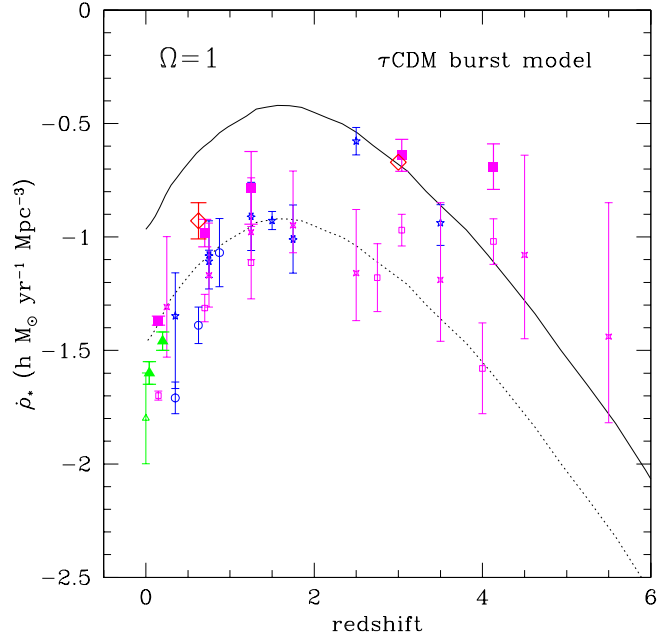,height=9truecm,width=9truecm}}
\caption{The star formation history in an $\Omega=1$ universe. Small open and large
filled points show the observations without and with (respectively)
corrections for dust extinction (see Appendix). The solid line shows a
collisional starburst model in a $\tau$CDM cosmology, normalized to
the Tully-Fisher relation at $z=0$.  The dotted line shows the same
model normalized to the present-day dust-corrected UV luminosity
density
\protect\cite{treyer:98}. }
\label{fig:taumadau}
\end{figure}
With improving observational constraints, the freedom to choose
cosmological parameters as one wishes is rapidly dissappearing, and
therefore we have chosen to focus on the popular
\lcdm\ cosmology. Any cosmology within the current observationally
favoured regime of parameter space, namely $\Omega=0.3-0.5$ (open or
flat) and $\Gamma
\simeq 0.2$, would give similar results. The old standard, cluster
normalized SCDM, would also give similar results because of the large
amount of power on small scales. Models with $\Omega=1$ and realistic
power spectra, such as $\tau$CDM, CHDM, or tCDM (see SP) show a
stronger peak at $z\sim1.5$ and a sharp decline in the star formation
rate at higher redshift. In Fig.~\ref{fig:taumadau} we show the star
formation history for a model in which we use the $\tau$CDM cosmology
defined in SP and normalize to the local Tully-Fisher relation in our
usual way.  This figure should be compared to the corresponding
Fig.~\ref{fig:new_madau_nodust} and \ref{fig:newmadau_dust} for our
standard \lcdm\ cosmology.  The $\tau$CDM model overproduces the
luminosity density at low redshift ($z\la2$) and falls below the
observations at high redshift ($z\sim 4$).  If we normalize the model
to fit the local UV-luminosity density, we find that there is not
enough star formation at high redshift  compared to the dust-corrected
new Madau diagram.  These models also violate the constraint on
$\Omega_{\rm gas}$ from DLAS at high redshift.  The results for models
with similar power spectra such as CHDM or tCDM would be nearly
identical.

\subsection{Stellar Population Synthesis}
\label{sec:variations:sed}
\begin{figure}
\centerline{\psfig{file=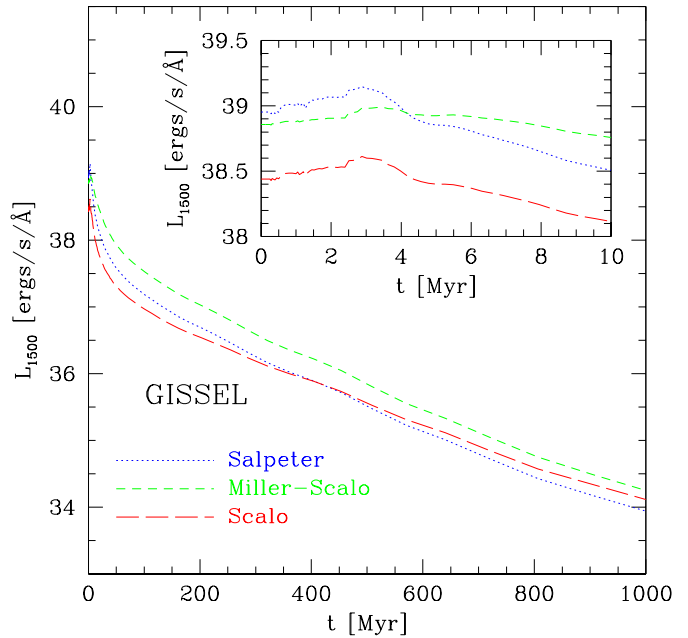,height=9truecm,width=9truecm}}
\caption{The luminosity of a single burst of mass $10^6$ \msun\ as a 
function of age for the GISSEL models with solar metallicity and
various IMF. Retaining the Salpeter shape but changing the lower mass
cutoff from 0.1 \msun\ to 1 \msun\ would result in shifting the
luminosities up by about a factor of three. }
\label{fig:imfcomp}
\end{figure}
\begin{figure}
\centerline{\psfig{file=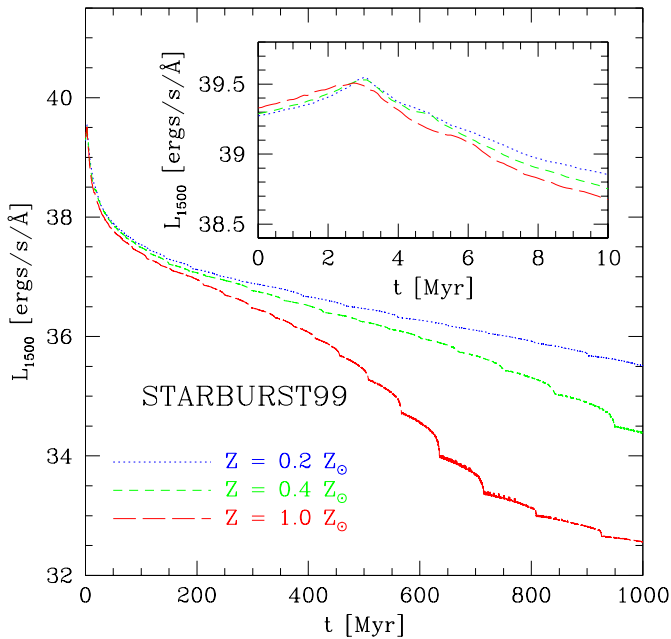,height=9truecm,width=9truecm}}
\caption{Same as Fig.~\protect\ref{fig:imfcomp}, but for the STARBURST99
models with different metallicities. The far-UV luminosity does not
depend much on the metallicity for the very young stars that dominate
it. }
\label{fig:metcomp}
\end{figure}

Most of the currently available observations of galaxies at $z \ga 3$
are viewing light that is emitted in the far UV ($\sim$ 1500
\AA). This light is dominated by very young ($\la 10$ Myr), massive O
and B type stars, which are notoriously poorly understood and
difficult to model.  In addition, the stellar mass-to-light ratio
depends on the metallicity and IMF.

We have made a comparison of the luminosity at 1500 \AA\ (averaged
over a tophat with width 400 \AA) of an instantaneous burst for a
variety of the stellar population synthesis models that are publically
available: GISSEL \cite{bc:93}, PEGASE \cite{pegase}, STARDUST
\cite{devriendt}, and STARBURST99
\cite{starburst99}
\footnote{The GISSEL
models were provided to us by S. Charlot, the PEGASE models by
M. Fioc, and the STARDUST models by J. Devriendt. The STARBURST99
models were downloaded from the public website
http://www.stsci.edu/science/starburst99/}. For models with solar
metallicity and a Salpeter IMF, all of these models agree within 0.15
dex for populations younger than 10 Myr old, which dominate the UV
light especially in starburst galaxies. The GISSEL, STARDUST, and
PEGASE models agree at a similar level for older populations as well,
but the STARBURST99 models differ from the others by as much as a
factor of fifteen for populations up to 1 Gyr in age.  We find a
similar level of agreement for populations with 0.2 and 0.4 solar
metallicity.

Fig.~\ref{fig:imfcomp} shows the dependence on the assumed IMF for the
solar metallicity GISSEL models. We see that the UV luminosity is
quite sensitive to the IMF, in particular to the fraction of high-mass
vs. low-mass stars. The Scalo IMF gives results that are lower by
about a factor of three at ages younger than 10 Myr. Similarly,
changing the lower- or upper-mass cutoff would have an effect on the
results. For example, if we 
drastically increased the lower-mass cutoff to 1 \msun\ instead of 0.1
\msun\ and retained the Salpeter shape, the curve would shift up by
about a factor of three
\footnote{We base this on a model that we ran using the STARBURST99 website}. 
None of these changes are huge in the context of the theory, but
evolutionary changes in the IMF as a function of redshift could induce
a tilt in the Madau diagram, and thus affect comparison with
observations.  However, to salvage the constant efficiency quiescent
model would require a total evolutionary change of a factor of 10 in
mass-to-light ratio (see Fig.~\ref{fig:newmadau_dust}), which would
stretch all the above effects to their limits.

The STARBURST99 models have been constructed to be particularly
relevant to the very young stellar populations that we are interested
in here, and use different stellar atmosphere models and a different
approach to modelling mass loss than the other three models mentioned
above (see
\citeNP{starburst99}).
In Fig.~\ref{fig:metcomp}, we show the dependence on metallicity in
these models. For models with metallicities ranging from one-fifth
solar to solar metallicity, the luminosity at 1500 \AA\ is almost
independent of metallicity for stars younger than 10 Myr. The
luminosity becomes more sensitive to metallicity for older stars ($\ga
200$ Myr).

Note that we have neglected the contribution from nebular emission in
our modelling. In the STARBURST99 models (Salpeter with solar
metallicity) nebular emission contributes about 15 percent to the
luminosity at 1500 \AA\ at an age of 1 Myr.  

\subsection{Dust and Reddening}
\label{sec:variations:dust}
Our treatment of dust and reddening is perhaps the least secure of our
assumptions.  This treatment falls into three broad categories: the
mean absorption per typical galaxy, the variation in absorption versus
wavelength (reddening curve), and the variation in absorption versus
galaxy luminosity (brighter galaxies are more extinguished).  None of
these threaten the two main conclusions of this paper, that the
collisional starburst model is consistent with present data on LBGs,
and that star formation efficiency at high redshift must be higher
than locally.  With regard to the first conclusion, we are claiming
only consistency using the best information presently available.  This
has certainly been shown.  With regard to the second point, even if we
reduce the the dust correction to zero, the constant efficiency
quiescent model is inconsistent with the \citeN{steidel:99} data at
$z\sim4$ and preliminary detections of $z\ga 5$ galaxies. 

There are hints that some aspects of our dust treatment are in fact in
error.  For example, our model galaxies are systematically 0.2 mag too
blue in V$-$I (Fig.~\ref{fig:colormag}).  Fixing this may necessitate
changing the reddening curve to a steeper one closer in shape to the
SMC curve; however, the overall amount of absorption would still
remain roughly the same (see figure).  Our third assumption (that
brighter galaxies are more extinguished) is based in part on precisely
the HDF galaxies we are modelling --- after correction for extinction,
is it precisely the reddest objects that are bolometrically brightest
\cite{meurer:99}.  

Opinion seems to be converging on our first and most crucial
assumption, the average absorption per typical galaxy.  Given the
observed V$-$I colors of LBG galaxies, it is inconceivable that they
have zero reddening.  At the other extreme, dust absorptions three
times larger than we have assumed would predict many more
sub-millimeter sources than seen \cite{meurer:99}.  Tweaking the mean
absorption within these boundaries would not greatly alter our
picture.

On the other hand, a serious change to our assumed absorption
treatment would be the discovery that mean absorption at a fixed
luminosity declines strongly with redshift (the reverse variation is
implausible).  This would introduce a strong bias into all tests
versus time, which is the most important probe of LBG models.  Present
data on the very distant Universe are simply too sparse to know for
sure whether distant sources are systematically less obscured.  Future
sub-millimeter observations are needed to pin down this last, very
important point.

\section{Summary and Conclusions}
\label{sec:conclusions}
The process of star formation is one of the largest uncertainties in
galaxy formation modelling. We have explored three different recipes
for star formation and the implications for observations of galaxies
at high redshift. We investigated a model in which quiescent star
formation has a constant efficiency, and galaxy-galaxy mergers trigger
bursts of star formation. In this model, we find that most of the
observable Lyman-break galaxies are collisional starbursts. We also
considered two models in which star formation occurs only in a
quiescent mode, either with constant or ``accelerated'' efficiency.

We find that our collisional starburst model produces excellent
agreement with the observed number density of bright galaxies as a
function of redshift from $2 \leq z \leq 6$ when a reasonable amount
of dust extinction is accounted for. The accelerated quiescent model
produces nearly the same number density of bright galaxies as the
starburst model over this redshift range. In contrast to the constant
efficiency quiescent model, which predict that very few bright
galaxies would be in place at $z\ga4$, the burst model and the
accelerated quiescent model make the tantalizing prediction that the
comoving number density of bright galaxies at redshift 5 and 6 is
nearly as high as at $z\sim3$. While this is consistent with the
numbers of galaxies in the HDF-N and S with photometric redshifts in
this range \cite{chen:99a,lanzetta:99a}, and some spectroscopic
confirmations of very high redshift galaxies exist, secure estimates
of the number densities of galaxies at these very high redshifts may
have to wait for NGST and results from the new generation of infrared
spectrographs on large ground-based telescopes.

A characteristic feature of the collisional starburst model is that
the \emph{unextinguished} rest-UV luminosity function at $z\sim3$ is
much flatter than a Schechter function at the bright end. Our
prediction of the intrinsic luminosity function is in excellent
agreement with the dust-corrected luminosity function of observed LBGs
calculated by
\citeN{adelberger:00}. Conversely, if we introduce the effects of dust
into the models using a simple, empirically motivated model of
differential dust extinction \cite{wh}, we obtain very good agreement
with the actual observed luminosity function. Both the accelerated
quiescent and the constant efficiency quiescent models show a steeper
decline at the bright end. The constant efficiency quiescent model
produces reasonable agreement with the observed bright end of the
luminosity function at $z\sim3$ \emph{if dust extinction is
neglected}, but with a realistic correction for dust extinction, these
models predict a cutoff on the bright end that is much too steep. The
accelerated quiescent model also has the wrong \emph{shape} compared
to the dust-corrected luminosity function of
\citeN{adelberger:00}. This model therefore underpredicts the number density 
of the very brightest LBGs (${\mathcal R}_{AB} < 22$ at $z\sim3$).

The global star formation history of the Universe is reflected in the
redshift evolution of the total densities of star formation, cold gas,
stars, and metals. We have compiled a new Madau diagram, using
recent observations and an observationally based dust correction. We
find a broad consistency between the dust-corrected optical estimates
of the star formation rate density and those obtained from far-IR
observations at $z\sim 0.7$ from ISO
\cite{flores:99} and sub-mm observations at $z\sim3$ from SCUBA
\cite{hughes:98}. The shape of the new Madau diagram is quite
different from the original one, with a more gradual rise in the star
formation rate from $z=0$ to $z\sim2$, and a plateau thereafter. The
shape of the Madau diagram produced by our models is quite sensitive
to our adopted recipe for star formation. The collisional starburst
model is in excellent agreement with the new dust-corrected data,
and the accelerated quiescent model is also in reasonable
agreement. The constant efficiency quiescent model shows a much
steeper decline with redshift, and have too little star formation at
high redshift compared with the new data.

Similarly, the collisional starburst model and the accelerated
quiescent model form their stars at a fairly high redshift, and thus
stellar ages are in agreement with observational constraints from
``fossil evidence'' contained in the ages of stars in nearby galaxies
\cite{renzini}. The constant efficiency quiescent model forms
its stars too late. In addition, we show that the redshift evolution
of the density of cold gas is consistent with constraints from DLAS
\cite{storrie-lombardi:96} in the collisional starburst model. 
The constant efficiency quiescent model may have too much cold gas
at high redshift, while the accelerated quiescent model may
have too little.

We investigate the predicted metal enrichment history of the Universe,
and find that an interesting side-effect of the burst and accelerated
quiescent scenarios is efficient pollution of the IGM and ICM at high
redshift due to supernovae-driven outflows. This may help to explain
the large covering factor implied by the high probability with which
C$_{\rm IV}$ and Mg$_{\rm II}$ absorption are seen in QSO spectra
whenever a bright galaxy at the appropriate redshift is $\la 30 \hkpc$
of the QSO line of sight
\cite{steidel:95}. It also implies that a substantial fraction of the 
metals at high redshift ($z\sim3$) are in the form of hot gas in
haloes or in the diffuse IGM, suggesting a solution to the ``missing
metals'' problem
\cite{pettini:99b,pagel}. When normalized to produce solar metallicity
stars in bright galaxies at redshift zero, with no introduction of
additional free parameters, the burst model simultaneously predicts
the correct observed metallicity of diffuse gas (Lyman-$\alpha$
forest) at $z\sim3$, as well as the observed metallicity of hot gas in
clusters at $z=0-0.3$. It is natural to associate the cold
interstellar gas in our model galaxies with the observed population of
DLAS. Here an interesting discrepancy is that the cold gas in our
models has a systematically higher metallicity than the DLAS, by about
a factor of $\sim 3$ at all redshifts. Moreover, as the constraints on
the metallicities of very low redshift ($z \la 1.0$) DLAS improve
\cite{pettini:lowzdlas}, if this gas is indeed the reservoir for star
formation, it becomes difficult to understand how it can become
enriched to solar values over such a short time interval. This favours
the explanation that the observed population of DLAS systematically
underestimates the metallicity of the total cold gas in galaxies.

The collisional starburst model produces good agreement with
observable individual properties of $z\sim3$ galaxies, such as the
half-light radii and internal velocity dispersions.  Our models also
predict that high redshift galaxies should have cold gas fractions
that are much larger than present day galaxies: values of $f_{\rm gas}
\equiv m_{\rm gas}/(m_{*}+m_{\rm gas})$ of 0.5 to 0.9 are typical. 
The model galaxies at $z\sim3$ have perhaps rather surprisingly high
metallicities, from 1/3 solar to solar for bright galaxies. The
stellar masses of the model LBGs range from $10^{8}-10^{11}$ h$^{-2}$
$\msun$, up to three orders of magnitude smaller than present-day
$L_{*}$ galaxies. If our models reflect the real Universe, this
indicates that the majority of the LBGs are not the fully-formed,
direct progenitors of today's $L_{*}$ galaxies, but must experience
considerable growth (via accretion or merging) by $z=0$ if they are
indeed the progenitors of massive present-day galaxies. Another
possible destiny for some of the smaller-mass LBGs ($\la 10^{9}$
h$^{-2}$ $\msun$) is that their stars, gas, and globular clusters are
stripped as they fall into the potential well of a nearby massive dark
matter halo, forming a Pop II stellar halo such as that of the Milky
Way
\cite{trager:97}. This resembles the picture of galactic
stellar halo formation outlined in the classic paper by
\citeN{searlezinn}, which is supported by much recent evidence
(see \citeNP{majewski:96}, \citeNP{carney:96},
\citeNP{sommerlarsen:97}, and references therein).

Finally, we compared our results with previous work and discussed the
effects that different model assumptions have on our results. In the
models of BCFL, star formation was made inefficient in small galaxies
by the combined properties of the recipes for star formation and
supernovae feedback. This led to the suppression of star formation at
high redshift, and a prediction of a strongly decreasing star
formation rate density with redshift. We argued that our constant
efficiency quiescent model gives overall very similar results to the
BCFL models, and therefore the same conclusions will apply: namely,
these models are not consistent with the recent data at $z\sim3$ and
$z\sim4$ when the observationally favored correction for dust
extinction is taken into account. The star formation recipe usually
used by Kauffmann et al. is very similar to our accelerated quiescent
model. It is interesting however that \citeN{kh} recently exchanged
this recipe in favor of one that is effectively rather similar to our
model with constant efficiency quiescent star formation plus
collisional starbursts. They disfavored the ``accelerated'' type model
both because they found, as we did, that too much gas was consumed to
be consistant with the observations of DLAS at $z\ga2$, but also
because it did not reproduce the observed redshift evolution of the
space density of bright quasars. Thus the same mechanism,
merger-driven inflows, may account for both high redshift galaxies and
quasars.

We showed that in cosmologies with $\Omega=1$ and realistic power
spectra (e.g., $\tau$CDM, CHDM, tilted CDM), the decline in the star
formation rate at high redshift is too steep to be consistant with the
dust-corrected data even in the burst model. However, in any cosmology
with parameters close to the values favored by a broad range of
observations ($\Omega_0=0.3-0.5$), we would have obtained similar
results to those presented here. We showed a comparison of the UV
stellar-mass-to-light ratio in a variety of different stellar
population models, and showed that there is reasonably good agreement
between the models produced by different groups. We showed that the UV
luminosites would have been about a factor of three lower had we
assumed a Scalo instead of Salpeter IMF, and about a factor of three
higher had we assumed a larger value for the lower mass cutoff (1
\msun\ instead of 0.1
\msun). We also showed that according to these models, 
the UV luminosity in galaxies with active star formation is nearly
independent of the metallicity of the stellar population.

We summarize our main conclusions as follows:
\begin{itemize}
\item The details of the recipes used to model star formation can have 
very large effects on the results of galaxy formation models (either
semi-analytic or numerical), especially when redshift evolution is
considered.

\item Models in which the efficiency of star formation is constant with 
redshift are strongly inconsistent with observations at high redshift
($z\ga3$) when the effects of dust extinction are taken into account,
and are inconsistent with higher redshift observations ($z\ga 4$) even
when dust is neglected.  Thus, a rather robust conclusion of this
study is that the efficiency of star formation (i.e. the star
formation rate per unit mass of cold gas in a galaxy) must increase
with redshift.

\item This increased efficiency can be accomplished in at least two 
physically plausible ways; either due to collisional starbursts or to
the scaling of the star formation rate with dynamical time or gas
surface density. Both models produce good agreement with most of the
observations considered here.

\item We favor the collisional starburst mechanism because it
gives better agreement with the shape of the $z\sim3$ luminosity
function at the very bright end, and it is in better agreement with
the constraints on the density of cold gas at high redshift from
DLAS. However, due to the remaining uncertainties in the modelling, we
do not consider the accelerated quiescent model to be strongly ruled
out.

\end{itemize}

The two successful models are based on completely different physical
ideas but are difficult to distinguish conclusively from the present
observations.  It is crucial to eventually determine which process is
actually the dominant mode of star formation at various redshifts.  In
one case, by studying high redshift galaxies we can expect to learn
something about the merger rate and the efficiency of merger-driven
inflows. In the other case, we expect to learn more about the masses
and internal properties of discs.

Both scenarios have observational support and theoretical motivation.
It may of course be that the true situation involves a combination of
both scenarios, however, simply adding starbursts to our usual
accelerated quiescent recipe does not provide a solution. This is
because the accelerated quiescent star formation very rapidly consumes
the cold gas supply at high redshift. The contribution from the burst
mode remains small because of the decreased gas fractions, and the
constraints on the cold gas at high redshift from DLAS are badly
violated. Real progress in this sort of modelling will only be made by
replacing some of our very simple recipes with more detailed and
physically motivated prescriptions. This will require a better
understanding of star formation in both normal and starburst galaxies.

An important observational aspect of the LBGs, which we have not
addressed at all in this paper, is their clustering
properties. Clustering in the collisional starburst scenario cannot be
accurately modelled with analytic approaches. Therefore we study this
using high-resolution N-body simulations in a companion paper
\cite{kolatt:99}. One might hope to discriminate between the 
collisional starburst and accelerated quiescent scenarios using the
clustering properties of galaxies, particularly the number of close
pairs. However, we show in a forthcoming paper \cite{spike2} that this
is not possible. We find that both scenarios are consistent with the
observed clustering properties at $z\sim3$.

There are, however, some direct observational tests which are feasible
in the near future and which may begin to discriminate between the two
scenarios. The collisional starburst scenario predicts that there
should be a relatively large population of bright, heavily
extinguished galaxies with star formation rates of hundreds of solar
masses per year. There are already preliminary indications from the
SCUBA results that this population has been detected. Future sub-mm
experiments with higher resolution and sensitivity will put stronger
constraints on the actual numbers and the associated total star
formation rates associated with these objects. Another way of
distinguishing the scenarios is from the morphologies of galaxies at
high redshift. In the collisional starburst picture, we expect many of
the LBGs to appear highly disturbed and to contain significant
substructure. In the accelerated quiescent scenario, we expect the
galaxies to appear smaller and denser, but otherwise similar to normal
local spirals. Again, there are preliminary indications that a large
fraction of the observed LBGs do show strong distortion and
sub-structure, even in rest visual bands
\cite{dickinson:nicmos,conselice:98}, but a more quantitative analysis
is needed. Finally, the observational sample of LBGs with measured
emission linewidths is growing rapidly. With better statistics and
more detailed modelling, these data will also help to discriminate
between the two scenarios.

\section*{Acknowledgements}
We thank Kurt Adelberger, Daniela Calzetti, Romeel Dav\'{e}, Mike
Fall, Mauro Giavalisco, Ken Freeman, Andrey Kravtsov, Max Pettini,
Jason Prochaska, Marcin Sawicki, and Chuck Steidel for useful
discussions.  We thank Carlton Baugh, Shaun Cole, Carlos Frenk, Cedric
Lacey, Guinevere Kauffmann, and Simon White for their feedback on
earlier versions of this work and clarifications of their models. We
also thank Caryl Gronwall, James Lowenthal, Jesus Gallego, Rafael
Guzman, Drew Phillips, and Scott Trager help with interpreting the
observations. We are grateful to Stephane Charlot for providing us
with the updated version of the GISSEL models. We thank Ken Lanzetta,
Hsiao-Wen Chen, and Amos Yahil for providing us with the Stonybrook
photometric redshift catalogs and for useful discussions of their
observations. We also thank Kurt Adelberger, David Hogg, James
Lowenthal, Richard McMahon, Lucia Pozzetti, Marcin Sawicki, Luc Simard
and Katherine Wu for providing observations in electronic form. RSS
acknowledges support from a GAANN fellowship at UCSC and a University
Fellowship from the Hebrew University, Jerusalem. This work was also
supported by a NASA ATP grant and NSF grant AST-9529098 at UCSC. We
also acknowledge the Institute for Theoretical Physics at UC Santa
Barbara, where this paper was completed.

\appendix
\section{How to Draw a Madau Diagram}
The star formation rate per unit comoving volume as a function of
redshift was first compiled from observations extending from redshift
zero to a redshift of about 4 by \citeN{madau:96}. The now-famous
``Madau diagram'' sketched out a picture of the history of star
formation, from a very early epoch when galaxies were perhaps first
forming until the present day. It has become popular to add more and
more points to this diagram. However, unfortunately, as different
authors have added their own points, the calculation of the derived
quantity, the \emph{total star formation rate density}, from what is
actually observed (luminosities of galaxies selected in some way) has
not always been consistent. We therefore think it is timely to revisit
the steps in calculating this derived quantity from the observations,
and to compile a set of points that have been calculated in as
consistent a way as possible. As most of the results quoted in the
literature assume an Einstein de Sitter cosmology, we also provide the
conversion to other cosmologies.

\begin{table}
\caption{Conversion factors from luminosity to star formation (SFR $= C \times L$), 
where SFR is the star formation rate in \msun yr$^{-1}$ and $L$ is the
luminosity in erg/s for H$\alpha$, and the luminosity density in
ergs/s/Hz for the other tracers \protect\cite{mpd:97}.}
\begin{tabular}{|cc|}
\hline
Tracer & conversion factor C \\
\hline
H$\alpha$ & $6.31 \times 10^{-40}$\\
L$_{1500}$ & $1.25 \times 10^{-28}$ \\
L$_{2800}$ & $1.27 \times 10^{-28}$\\
\hline
\end{tabular}
\label{tab:sfconv}
\end{table}

\begin{table*}
\caption{Incompleteness corrected luminosity densities at various redshifts, using various 
tracers of star formation, for $\Omega=1$ ($q_0=0.5$; EDS), 
$\Omega_0=0.3$, $\Omega_{\Lambda}=0.7$ (\lcdm) and $\Omega_0=0.3$, 
$\Omega_{\Lambda}=0.0$ (OCDM). Units are h$^{-2}$ erg s$^{-1}$ Hz$^{-1}$.}
\begin{tabular}{|ccccccc|}
\hline
Reference & tracer & redshift & $\log [\rho^{EDS}_{L}]$ & $\log [\rho^{\lcdm}_{L}]$
& $\log [\rho^{\rm OCDM}_{L}]$ & log error \\
\hline
\protect\citeNP{gallego} & H$\alpha$ & 0.0 & -1.80 & -1.80 & -1.80 & 0.2\\
\protect\citeNP{kiss} & H$\alpha$ & 0.0425 & -1.60 & -1.62 & -1.61 & 0.05 \\
\protect\citeNP{treyer:98} & $L_{2000}$ & 0.15 & -1.70 & -1.76 & -1.72 & 0.02 \\
\protect\citeNP{tresse:98} & H$\alpha$ & 0.2 & -1.46 & -1.53 & -1.49 & 0.04\\
\protect\citeNP{cfrs:sf} & $L_{2800}$ & 0.35 & -1.71 & -1.83 & -1.75 & 0.07\\
			&& 0.625 & -1.39 & -1.56 & -1.46 & 0.08\\ &&
			0.875 & -1.07 & -1.26 & -1.15 & 0.15\\
\protect\citeNP{flores:99} & $L_{\rm FIR}$ & 0.625 & -0.93 & -1.10 & -0.98 & 0.08 \\
\protect\citeNP{cowie} &$L_{2000}$ & 0.7 & -1.31 & -1.49 & -1.39 & 0.06\\
			&& 1.25 & -1.11 & -1.33 & -1.22 & 0.15\\   
\protect\citeNP{connolly} & $L_{2800}$ & 0.75 & -1.08 & -1.26 & -1.16 & 0.15\\
			&& 1.25 & -0.91 & -1.13 & -1.02 & 0.15 \\
			&& 1.75 & -1.01 & -1.25 & -1.14 & 0.15 \\ 
\protect\citeNP{madau:96} & $L_{1500}$ & 2.75 & -1.4 & -1.65 & -1.56 & 0.15\\
			&& 4.0 & -1.90 & -2.16 & -2.08 & 0.2\\
\protect\citeNP{mpd:97} &$L_{1500}$& 2.75 & -1.18 & -1.43 & -1.34 & 0.15 \\ 
			&& 4.0 & -1.58 & -1.84 & -2.08 & 0.2 \\
\protect\citeNP{steidel:99} & $L_{1500}$ & 3.04 & -0.97 & -1.22 & -1.13 & 0.07\\
			&& 4.13 & -1.02 & -1.27 & -1.2 & 0.1\\
\protect\citeNP{hughes:98} & sub-mm & 3.0 & -0.67 & -0.92 & -0.83 & 0.16\\
\protect\citeNP{sly} & $L_{2800}$ & 0.35 & -1.34 & -1.46 & -1.39 & (+0.19)(-0.32)\\
		&& 0.75 & -1.11 & -1.29 & -1.19 & (+0.05)(-0.07)\\
		&& 1.5 & -0.93 & -1.16 & -1.05 & 0.04\\
		&& 2.5 & -0.58 & -0.83 & -0.73 & 0.06\\
		&& 3.5 & -0.94 & -1.19 & -1.10 & (+0.08)(-0.1)\\
\protect\citeNP{plf} & $L_{1500}$ & 0.25 & -1.31 & -1.40 & -1.34 & (+0.31)(-0.22)\\
			&& 0.75 & -1.17 & -1.35 & -1.25 & (+0.23)(-0.14)\\
			&& 1.25 & -0.98 & -1.20 & -1.09 & (+0.24)(-0.12) \\
			&& 1.75 & -0.95 & -1.19 & -1.08 & (+0.24)(-0.12)\\
			&& 2.5 & -1.16 & -1.41 & -1.31 & (+0.28)(-0.21)\\
			&& 3.5 & -1.19 & -1.44 & -1.36 & (+0.34)(-0.27)\\
			&& 4.5 & -1.08 & -1.34 & -1.26 & (+0.44)(-0.37)\\
			&& 5.5 & -1.44 & -1.70 & -1.63 & (+0.59)(-0.38)\\ 
\hline
\end{tabular}
\label{tab:madau}
\end{table*}

To draw your own Madau plot, follow these steps:
\begin{enumerate} 
\item {\bf Correct for Incompleteness}

Observational samples are generally flux limited, and thus the
intrinsic luminosity of the faintest objects in the sample changes
with redshift. In order to understand the true redshift dependence of
the \emph{total} luminosity density, one must correct the observations
for incompleteness. This is most easily done by fitting a functional
form (i.e. a Schechter function) to the luminosity function obtained
from the observations themselves. If the usual parameters of the
Schechter function, $\phi_{*}$, $L_{*}$, and $\alpha$, are given, then
the total luminosity density is given by $\phi_{*} L_{*}
\Gamma(2+\alpha)$. Here $\Gamma$ is the usual Gamma function 
\cite{crc}. Note
that for values of $\alpha$ steeper than $-1$, faint galaxies
contribute a substantial fraction of the total luminosity density and
therefore the results are quite sensitive to the faint end slope,
which is often poorly constrained. This step leads to the first source
of inconsistency: different authors have assumed different values of
$\alpha$, or have integrated down to different lower limiting
luminosities.

\item {\bf Convert to the Desired Cosmology}

Most observational references quote luminosity densities assuming an
Einstein de Sitter ($\Omega=1$) cosmology. To convert from one
cosmology to another, one must take into account two effects. The
luminosity derived from a given apparent magnitude will change, as
will the comoving volume derived from a given angular size and
redshift range. Luminosities scale as
\begin{equation}
\log \left(\frac{L_{\rm new}}{L_{\rm old}}\right) = 
 2 \log \left(\frac{d^{\rm old}_L(z)}{d^{\rm new}_{L}(z)}\right) \, ,
\end{equation}
where $L_{\rm old/new}$ and $d^{\rm old/new}_{L}(z)$ are the
luminosity and luminosity distance at a given redshift in the old and
new cosmologies. If we now define $f_{V} \equiv V_{\rm new}/V_{\rm
old}$ as the ratio of the comoving volume in the new cosmology to that
in the old cosmology, then the luminosity density scales as
\begin{equation}
\log \left(\frac{\rho^{\rm new}_{L}}{\rho^{\rm old}_{L}}\right) =
2 \log \left(\frac{d^{\rm old}_L(z)}{d^{\rm new}_{L}(z)}\right) -
\log[{f_V(z)}]
\, .
\end{equation}

\item {\bf Convert Luminosity to Star Formation}

The usual tracers of star formation are the luminosity of nebular
emission lines like H$\alpha$ or O$_{\rm II}$, or the far-UV
(1500-2800 \AA) continuum. This observable quantity must be converted
into a star formation rate. This conversion generally relies on
stellar population models and an assumed star formation history and
IMF (see Section~\ref{sec:variations:sed} for a discussion of the
attendant uncertainties and a quantitative comparison of different
stellar population models).  We give a compilation of conversion
factors for various tracers of star formation in
Table~\ref{tab:sfconv}. These are taken from \citeN{mpd:97}, assuming
a Salpeter IMF. The use of different conversion factors is a second
source of inconsistency in published results in the literature.

\item {\bf Correct for Dust Extinction}

If the tracer of star formation is an optical or UV luminosity, then
the effects of dust extinction may be non-negligible. In the original
Madau plot, no attempt was made to correct for extinction due to the
lack of knowledge at that time about the effect that dust was likely
to have on the observations, particularly at high redshift. As we
discussed in the main text (see Section~\ref{sec:models:dust}), some
observational estimates of the extinction in the UV ($\sim$1500--2000
\AA) are now available. These seem to indicate that the amount of
extinction at this wavelength in nearby starburst galaxies is similar
to that in the $z\sim 3$ LBGs, if the correlation between
spectral slope and dust extinction remains the same. However, the
amount of extinction depends strongly on luminosity, with
instrinsically brighter (more rapidly star-forming) galaxies being
more heavily extinguished. Therefore it is probably not a good idea to
apply a fixed correction factor to galaxies of all luminosities, as we
would in effect be doing if we first corrected the luminosity density
for incompleteness, as described in step 1, and then multiplied by a
fixed factor to correct for dust extinction. If this correction factor
was derived from observations of bright galaxies, but a substantial
fraction of the total luminosity density is contributed by faint
galaxies, this will lead to an overestimate of the total,
dust-corrected luminosity density.
\end{enumerate}

We give a compilation of star formation densities derived from various
tracers at various redshifts in Table~\ref{tab:madau}, for three
different cosmologies (see Table caption). These values have been
corrected for incompleteness by integrating over the entire luminosity
function and converted from luminosity to star formation rate using
the conversion factors from Table~\ref{tab:sfconv}, but no correction
for dust extinction has been made. Fig.~\ref{fig:new_madau_nodust}
shows the resulting Madau plot for the \lcdm\ cosmology, and
Fig.~\ref{fig:taumadau} for an $\Omega_0=1$ cosmology.

We now select a subset of these observations, those which we believe
to give the most robust estimates of the star formation rate density
over a broad range of redshifts. These samples represent a reasonably
large volume (unlike the results from the HDF), are based on
spectroscopic redshifts (unlike the results based on the more
uncertain photometric redshifts), and were observed in roughly the
same rest waveband (1500-2000 \AA), and so do not require large
photometric extrapolations or interpolations. We calculate the
integrated UV luminosity density at low redshift from the results of
\citeN{treyer:98}, from \citeN{cowie:99} at intermediate redshift, 
and from \citeN{steidel:99} at high redshift. The sample with the best
constrained faint-end slope is the lowest redshift sample of
\citeN{treyer:98}, which has a derived slope $\alpha=-1.6$, exactly
the same as the faint end slope at $z=3$ derived from the combined
ground-based and HDF samples of LBGs
\cite{steidel:99}. We see no particular reason that the UV luminosity
function should flatten at intermediate redshifts, so we assume
$\alpha=-1.6$ for the Cowie et al. sample as well, and calculate all
the incompleteness corrections accordingly (we use the values of
$\phi_{*}$ and $L_{*}$ quoted by \citeN{cowie:99} for a fixed
$\alpha=-1.5$, but we actually use $\alpha=-1.6$ in computing the
luminosity density).

The results of \citeN{steidel:99} indicate that there is a factor of
4.7 extinction at $\sim 1500$ \AA\ in LBGs brighter than ${\mathcal
R}=25.5$, which corresponds to $0.4 L_{*}$ at the mean redshift of the
sample ($z\sim3$) and for our assumed cosmology. We will assume that
these results hold for all of the UV-selected samples at all
redshifts.  To obtain a conservative (``minimal'') dust correction, we
apply the factor of 4.7 correction to all galaxies brighter than $0.4
L_{*}$ (using the appropriate value of $L_*$ at each redshift) and no
correction for galaxies fainter that this limit, for each of the three
samples (\citeNP{treyer:98}, \citeNP{cowie:99}, and
\citeNP{steidel:99}). Note that according to our previous calculation of dust 
extinction as a function of luminosity using the Wang \& Heckman
scaling, this will underestimate the dust correction but not by much
(see Fig.~\ref{fig:lf_z3_burst}). Because we have assumed that the slope of
the UV luminosity function is the same for all of these samples,
this is equivalent to adding 0.33 in the log to the values tabulated
in Table~\ref{tab:madau}.  To obtain a ``maximal'' dust correction, we
apply the factor of 4.7 to
\emph{all} galaxies in the sample. As we discussed in item 4 above,
this is likely to be an overestimate.
Note too that if we integrate the new Madau diagram with the
\emph{fiducial} dust correction, the total mass of stars produced by
$z=0$ is in good agreement with the estimates of \citeN{fhp}. However,
if we integrate the Madau diagram with the
\emph{maximal} dust correction, the total mass in stars is a factor 
of five too large (see Fig.~\ref{fig:omega_star}).  This suggests that
something in between the minimal and maximal dust correction is
probably reasonable (though one could get around this conclusion
by varying the the IMF from the assumed Salpeter shape).

We plot the results for our fiducial cosmology in
Fig.~\ref{fig:newmadau_dust}. For comparison we also plot the
results at low redshift from recent H-$\alpha$ surveys, corrected for
incompleteness and dust extinction by the original authors
\cite{kiss,tresse:98}. These surveys are quite deep, so incompleteness
corrections are less important, and because H$\alpha$ is emitted at
$\sim$6500\AA, the effects of dust are also less severe. These results
agree well with the \citeN{treyer:98} point. There is also a pleasing
consistency between the dust-corrected, UV-based results at
intermediate redshift and the Far-IR results based on ISO observations
at $z\sim0.7$
\cite{flores:99}, and finally the high redshift UV-based results and 
the sub-mm results from SCUBA observations at $z\sim3$
\cite{hughes:98}\footnote{Though it should be noted that 
deriving the star formation rate from the IR and sub-mm observations
is a more complex operation. See the discussions and more detailed
modelling by \protect\citeN{trentham:99} and
\protect\citeN{blain:99}}.

A very different star formation history emerges from this ``new''
Madau plot. Instead of the steep rise from $z=0$ to $z\sim1.5$, there
is a more gradual rise, and instead of the peak at $z\sim2$ and
fall-off at higher redshift, there is a plateau.

\bibliographystyle{mnras}
\bibliography{mnrasmnemonic,spf}
\end{document}